\documentclass[12pt]{article}

\usepackage{sbc-template}

\usepackage{graphicx,url}

\usepackage{graphicx} 
\usepackage{subfigure} 
\usepackage{algorithm}
\usepackage[noend]{algpseudocode}
\usepackage{tikz}
\usetikzlibrary{positioning}
\usepackage{booktabs}
\usepackage{latexsym}

\usepackage[utf8]{inputenc}

\title{Fast Reroute with Highly Connected Routes Based on \\ Maximum Flow Evaluation}

\author{Leon Okida\inst{1}, Maverson E. S. Rosa\inst{1}, Elias P. Duarte Jr.\inst{1}}

\address{Dept. Informatics - Universidade Federal do Paraná (UFPR)\\
  Curitiba -- Brazil
  \email{\{laog19,maverson,elias\}@inf.ufpr.br}
}

\begin{document}

\maketitle

\begin{abstract}
Fault-tolerant routing allows the selection of alternative routes to the destination after the route being used fails. Fast Reroute (FRR) is a proactive strategy through which the protocol pre-configures backup routes that are activated when needed. In this work, we propose the MaxFlowRouting algorithm that employs maximum flow evaluation as well as the route size to select routes that are highly connected. The main advantage of the proposed algorithm is that if any component of such a route fails, there are more alternative paths to the destination in comparison with the route computed with Dijkstra’s shortest path algorithm. Simulation results are presented in which we compare the two algorithms (Dijkstra’s and MaxFlowRouting) for multiple different random graphs (including Erdos-Renyi, Barábasi-Albert, and Watts-Strogatz) and also for the topologies of some of the most important Internet backbones of the U.S.A., Europe, Brazil, and Japan: Internet2, Géant, RNP, and Wide.
\end{abstract}

\section{Introduction}

Organizations and individuals have become increasingly dependent on computer networks, and the Internet in particular. The Internet can present highly unstable behavior at some instances \cite{duarte2010finding}. A robust network must adopt fault-tolerant routing strategies in order to recover quickly after the occurrence of failures along the network infrastructure. Fast Reroute (FRR) \cite{shand2010ipfrr} is a proactive strategy that activates a backup route when the primary route fails. FRR has been adopted by several Internet protocols, including IP itself \cite{shand2010ipfrr}, MPLS \cite{pan2005fast,duarte2012gigamanp2p}, and OSPF \cite{filsfils2012loop}.  The alternative to FRR is to employ a reactive strategy, which causes the disruption of traffic delivery after some failure of a link or network device until the tables re-converge to the new topology. Packets for destinations that were previously reached through the failed route will most certainly be dropped during that re-convergence period, which can cause serious disruptions of distributed services and applications.

FRR computes backup routes that are stored in the routing tables as alternatives. In this way, after a router detects a route failure, it can immediately try an alternative path to the destination. In this case, the disruption time is limited by how long it takes to detect the failure and invoke the backup route. One must note, however, that the success of FRR highly depends on the existence of alternative routes that can effectively bypass the failed router or link. 

In this work, we propose the MaxFlowRouting algorithm based on maximum flow evaluation \cite{ffulk, cormen} to select robust routes. Maximum flow evaluation implicitly finds routes that are as ``connected'' as possible, given the network topology. Thus it takes into consideration the redundancy of routes available, since the greater the maximum flow, the greater the number of disjoint paths and, consequently, the greater the number of alternative backup routes that can be used in case of a failure.

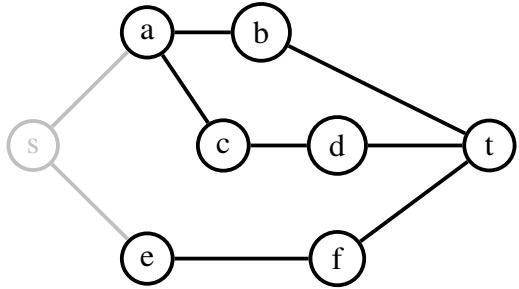
\begin{figure}[htb]
\centering
\begin{tikzpicture}[
    vertex/.style={circle, draw, thick, line width=0.5mm, minimum size=0.6cm},
    grayvertex/.style={vertex, lightgray},
    edge/.style={-, line width=0.5mm},
    grayedge/.style={edge, lightgray}
    ]

  \node[grayvertex] (s) at (0, 0) {s};
  \node[vertex] (a) at (1.5, 1.5) {a};
  \node[vertex] (e) at (1.5, -1.5) {e};
  \node[vertex] (b) at (3, 1.5) {b};
  \node[vertex] (c) at (2.5, 0) {c};
  \node[vertex] (f) at (4, -1.5) {f};
  \node[vertex] (d) at (4, 0) {d};
  \node[vertex] (t) at (6, 0) {t};

  \draw[grayedge] (s) -- (a);
  \draw[grayedge] (s) -- (e);

  \draw[edge] (e) -- (f);
  \draw[edge] (a) -- (b);
  \draw[edge] (a) -- (c);
  \draw[edge] (c) -- (d);
  \draw[edge] (b) -- (t);
  \draw[edge] (d) -- (t);
  \draw[edge] (f) -- (t);

\end{tikzpicture}
\caption{Example of next hop selection based on maximum flow evaluation and route size.}
\label{fig:cap_criterios_ex2}
\end{figure}

MaxFlowRouting builds a table with multiple next-hop alternatives for each destination. Routes are ordered according to two criteria: maximum flow and route size. Figure \ref{fig:cap_criterios_ex2} shows an example of how a router running MaxFlowRouting orders the next hops to a given destination. In the figure, $s$ is the source and $t$ is the destination.  The algorithm is being executed by node $s$. There are two alternative adjacent links to the destination: $(s,a)$ and $(s,e)$. MaxFlowRouting evaluates the two alternatives by first removing the source node (that is running the algorithm) plus its adjacent edges from the graph. The subgraph generated is formed by the darker lines.

To evaluate edge $(s,a)$, node $s$ computes the maximum flow and shortest path criteria from node $a$ to node $t$ (the destination) and from $e$ to $t$. The route from  $a$ yields a maximum flow of 2, while the route from $e$ yields a maximum flow of 1. This indicates that there are two disjoint paths from node $a$ to node $t$, but a single one from node $e$ to node $t$. Furthermore, the route size is also computed, being equal to 2 from $a$ to node $t$ as well as from $e$ to $t$. Weights are used to define the influence of the two criteria on route selection. Since the focus of the algorithm is on fault tolerance, a relatively higher weight should be employed for max-flow. In the example, the route through $a$ has a preference over the route through $e$ due to the higher flow value.

We also describe an FRR algorithm based on {\em backtracking} that works on any routing table with multiple alternatives per destination: when one alternative route fails, the packet can return to try another alternative. Thus the routing succeeds even if the routing table maintained by the node running the algorithm does not reflect the latest changes to the network. Note that the routes can be produced by MaxFlowRouting or any other routing algorithm, such as Dijkstra's. We show this algorithm works even if a single unknown route is fully functional.

Simulation results are also presented comparing MaxFlowRouting with Dijkstra's shortest path algorithm. The two algorithms (Dijkstra’s and MaxFlowRouting) were executed on multiple different topologies, such as random graphs (including Erdos-Renyi, Barábasi-Albert, and Watts-Strogatz) and also on some of the most important Internet backbones of Brazil, the USA, Europe, and Japan: RNP, Internet2, Géant, and Wide. Results confirm that MaxFlowRouting provides more backup routes to the destination with slightly larger routes than those of Dijkstra's algorithm.

The rest of this work is organized as follows. Section 2 presents MaxFlowRouting and the FRR algorithm based on backtracking. Section 3 presents experimental results obtained with simulation. Related work is in Section 4, and the conclusions in Section 5.

\section{The MaxFlowRouting Algorithm}
\label{cha:proposta}

This section presents the proposed MaxFlowRouting algorithm. Furthermore, an FRR algorithm with backtracking is also presented,  Note that both algorithms are independent: MaxFlowRouting can produce routes that are employed by another FRR algorithm, and the FRR with backtracking algorithm can use routes computed by any routing algorithm. After the specification of the algorithms, case studies are presented, and the proof that the FRR algorithm successfully routes a message to the destination even if there is a single fully functional route and the routing table is not up-to-date with the real network topology.

\subsection{MaxFlowRouting: Specification}
\label{ssec:prop_especificacao}

A node running the algorithm maintains the graph $G=(V,E)$ locally as a representation of the network topology. Note that algorithm does not require the graph to exactly reflect the real network topology, after any node detects some topology change, this should be disseminated to the other nodes. The routing table is updated after a node receives the new topology information and updates the graph $G$ accordingly. Each node runs Algorithm \ref{fig:PseudoCodeGenRoutingTable} (\texttt{MaxFlowRouting}) to generate and update the routing table for all possible destinations. As described next, \texttt{MaxFlowRouting} uses a combination of maximum flow evaluation and route size to order the candidate next hops for each destination. 

Maximum flow evaluation \cite{cormen,schroeder2004computing} allows the determination of the path of greatest capacity between two vertices of a graph $G$. In this work all edge capacities are the same and equal to 1. Actually, the maximum cut is equivalent to another abstraction whose application to fault-tolerant routing is even more intuitive: that is the minimum cut, defined next. The minimum cut $m$ between two vertices $u$ and $v$ of graph $G$ is a set of edges $C$ such that, the removal of all edges in $C$ from $G$ disconnects $u$ and $v$, i.e. there is no path in $G$ from $u$ to $v$. The size (or cardinality) of a cut $C$, denoted by $|C|$, is the number of edges in $C$. Furthermore, a cut $C$ is called the \textit{minimum cut} if, for every cut $C'$ between the same pair of nodes, $|C| \leq |C'|$. For any pair of vertices in the network, the maximum flow and the minimum cut have the same cardinality and can be computed using the same algorithms.

The selection of the next hop is done through the evaluation of all neighbors. An evaluation function $\Gamma(G',i)$ is executed by each node $s$ for each of its neighbors $i$, on a graph  $G'=(V', E')$, where $V'= V - {s}$ and $E'= E - (s, j), \forall j | (s,j) \in E$. Function $\Gamma(G',i)$ returns a numeric value for each neighbor $i$ adjacent to $s$, the highest the better. 

Besides the maximum flow, $\Gamma(G',e)$ is computed also using the route size, so that it is calculated with the expression $\Gamma(G',e) = w_1 * c_1(e)  +  w_2 * c_2(e)$. In this expression, maximum flow evaluation (criterion $c_1$) and route size (criterion $c_2$) are the two criteria used to evaluate neighbor $i$ as the next hop to the destination. Each criterion has a weight ($w_1$ and $w_2$, respectively) which dictates how much each criterion influences the final value function $\Gamma(G',i)$ computes. Thus, from a given vertex $s$, all adjacent edges are sorted as the next hop of choice according to $\Gamma(G',i)$. Next hops with the highest values for $\Gamma(G',e)$ are chosen first for routing a message to a certain destination.

Algorithm \texttt{MaxFlowRouting} is executed by each node $i$, which initially removes itself from the $G$ known topology, generating the graph $G'$. After that, $s$ computes the maximum flow from each neighbor to the destination $t$.  Furthermore, $s$ computes the distance from each neighbor to each destination. Finally function $\Gamma(G', j)$ is computed for each neighbor to each destination, using the weighted maximum flow and distance. For each destination (routing table entry), there is a list of \textit{CandidateNextHops}. This ordered list is initialized empty. A neighbor is included in \textit{CandidateNextHops} if it has a path to the final destination in $G$. Finally, the edges in \textit{CandidateNextHops} are ordered according to the evaluation function $\Gamma(G', i)$.

\begin{algorithm}
\caption{Specification of the \texttt{MaxFlowRouting} routing algorithm: generation of the routing table by node $i$.}
\label{fig:PseudoCodeGenRoutingTable}
\begin{algorithmic}[1]
\Procedure{MaxFlowRouting}{s}
\State \textbf{let} $G' \gets G - \{i\}$
\For{each destination $t$ in $G'$}
    \State \textbf{let} $CandidateNextHops \gets$ empty list
    \For{each neighbor $j$ adjacent to $i$ in $G$}
        \If{there is no path through $j$ to $t$ in $G'$}
            \State \textbf{ignore} $j$
        \Else
            \State \textbf{add} $j$ to $CandidateNextHops$
            Compute $\Gamma(G', j) \gets MaxFlow(j,t)*w_1 + Distance(j,t)*w_2$
        \EndIf
    \EndFor
    \If{$CandidateNextHops$ contains more than one edge}
        \State $\forall j$ in $CandidateNextHops$ \textbf{sort} those neighbors according to function $\Gamma(G', j)$
    \EndIf
    \State $RoutingTable[t] \gets CandidateNextHops$
\EndFor
\EndProcedure
\end{algorithmic}
\end{algorithm}

The routing table generated with algorithm \texttt{MaxFlowRouting} is used as the basis for the message forwarding procedure such as the FRR with backtracking algorithm presented as Algorithm \ref{fig:PseudoCodePacketForwarding}. This algorithm uses the routing table to select the next hop to a given destination. 

Algorithm \texttt{FastReRouteWithBacktracking} is executed by node $i$ to forward a message to a destination node $t$. First of all, node $i$ checks whether it is a neighbor of node $t$, and in this case sends the message directly to $t$. Each message $msg$ carries the sequential list of vertices through which it has passed so far, called $msg.Visited$. Node $i$ adds itself to $msg.Visited$ in case it is not already there. If it is there and it receives the message again then a cycle is detected, and this can only happen after a message is backtracked from a neighbor that did not have any route to the destination. Now node $i$ selects the next hop to forward the message. It chooses the best-ranked next hop in the corresponding routing table entry which is not in $msg.Visited$. In case there is no such a node to forward the message to, node $i$ backtracks the message to node $h$, from which it had received the message. In case there is no $h$ before $i$ then $i$ is the source $s$ and there is no route from $s$ to $t$.

\begin{algorithm}
\caption{Specification of a Fast-ReRoute algorithm with backtracking.}
\label{fig:PseudoCodePacketForwarding}
\begin{algorithmic}[2]
\Procedure{FastReRouteWithBacktracking}{i}
\If{$t$ is adjacent to $i \in G'$}
    \State \textbf{send} message $msg$ to $t$
\ElsIf{$msg.Visited$ does not contain node $i$}
    \State \textbf{add} node $i$ to $msg.Visited$
\If{$RoutingTable[t]$ contains nodes that are not in $msg.Visited$}
    \State \textbf{choose} the following sorted next hop $j$ in $RoutingTable[t]$, such that $j \notin msg.Visited$
        \State \textbf{send} message $msg$ to $j$
    \Else // There is no node to forward the message to
        \If{$msg.Visited$ contains at least one node $h$ before $i$}
            \State \textbf{send} $msg$ back to $h$
        \Else // The message was backtracked to the source $s$
            \State \textbf{return} Error: There is no route from the source $s$ to $t$
        \EndIf
    \EndIf
\EndIf
\EndProcedure
\end{algorithmic}
\end{algorithm}

\subsection{MaxFlowRouting \& FRR with Backtracking: Case Studies}
\label{sec:prop_exemplificacao}

\begin{figure}
    \centering
    \subfigure[A link fails as a message is being transmitted.]{\label{fig:prop_rede1}\includegraphics[scale=.25]{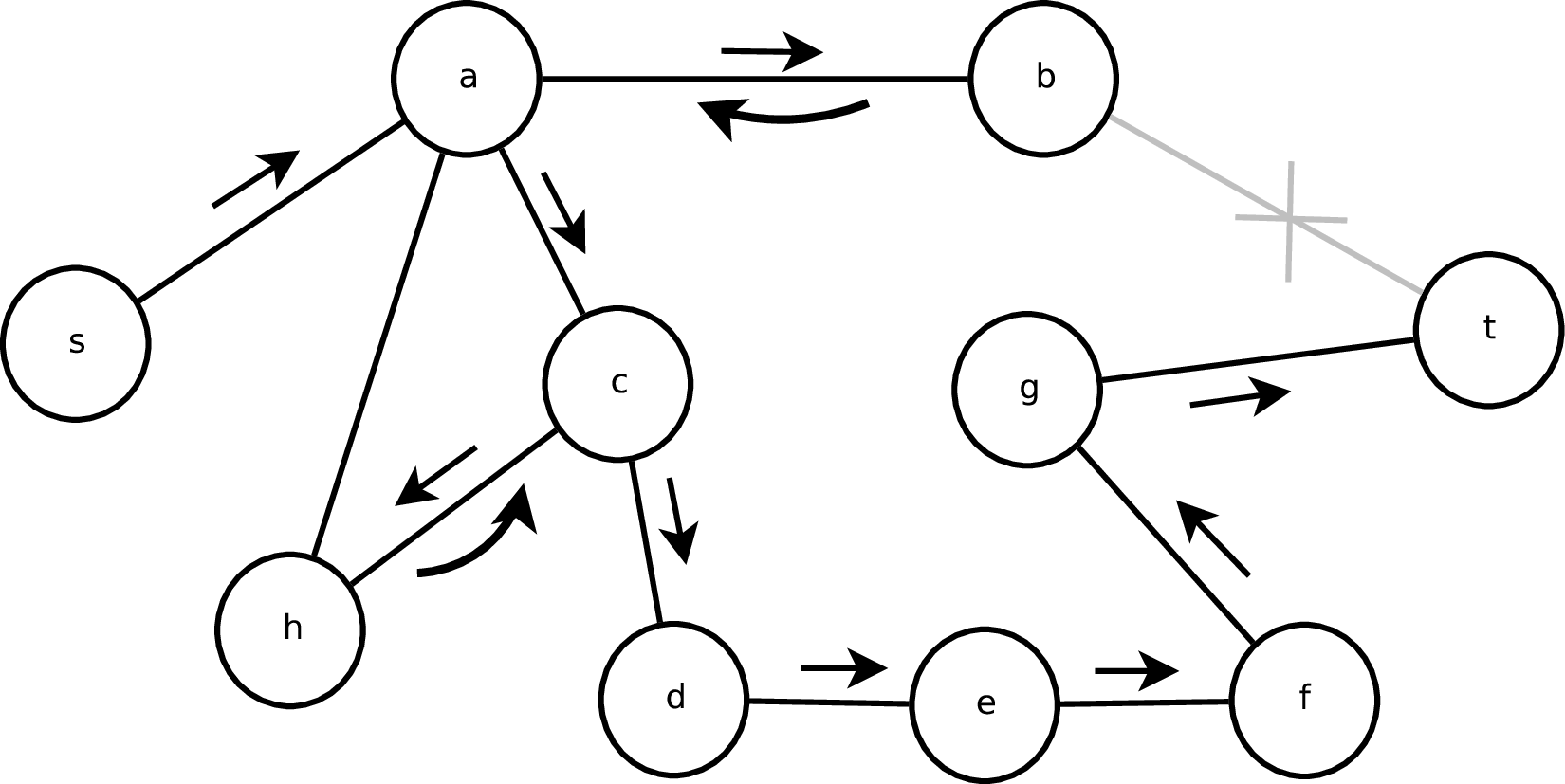}}
    \subfigure[The choice of an edge of a minimum cut.]{\label{fig:prop_rede2}\includegraphics[scale=.25]{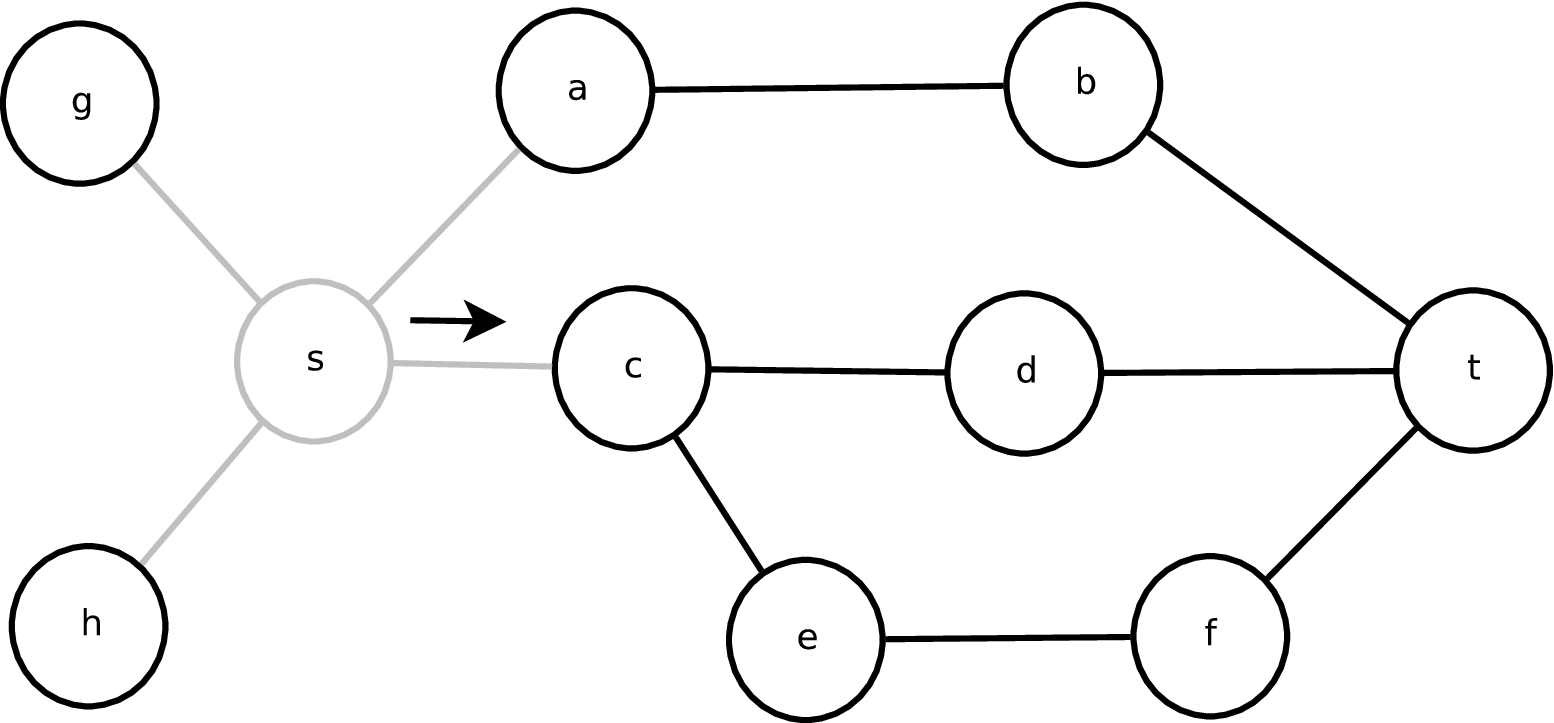}}
    \caption{MaxFlowRouting and FRR with Backtracking case studies.}
    \label{fig:prop_rede1_}
\end{figure}

Figures \ref{fig:prop_rede1} and \ref{fig:prop_rede2} show two case studies of the proposed MaxFlow Routing and FRR with backtracking algorithms. Figure \ref{fig:prop_rede1} presents a situation in which link $(b,t)$ fails. Node $b$ would use that link for forwarding a message. At the time the message is sent by the node $s$, only node $b$ is aware of the failure. Initially, node $s$ sends the message to node $a$, as it is its only routing alternative to the destination $t$. Considering that node $a$'s routing table presents node $b$ as the alternative that has the highest value returned by the evaluating function $Gamma$, $a$ forwards the message to node $b$, which does not have a route to the destination and thus ends up returning the message to the node from which it was received, i.e. node $a$. After receiving the same message again, node $a$ checks the routing table searching for the entry that has the highest value returned by the evaluation function $Gamma$ and which has not been previously visited by the message: the result is node $c$. Node $a$ then forwards the message to node $c$. At this point the main path taken by the message is represented as $s -> a -> c$, with node $b$ also marked as visited, but not included in the route actually used to reach the destination.

The routing table of node $c$ presents three alternatives $a, h, d$ (ordered according to function $\Gamma$) for the destination $t$. Node $c$ would thus first choose node $a$ however it is already marked as visited by the message. Thus node $c$ selects node $h$. Note that evaluation of node $h$ presents a higher value than that of node $d$, because, when node $c$ and all its adjacent edges are removed from graph $G'$, despite both nodes $h,d$ having the same value for the maximum flow, the minimum path to node $t$ is shorter through $h$ ($h -> a ->b -> t$), node $c$ considers that path as it is unaware that link $(b,t)$ has failed. As it was in $b$, When node $h$ receives the message, it learns that there are no working routes for the destination, and node $b$ has been already visited by the message. Thus node $h$ returns the message to node $c$, which now forwards the to node $d$. From node $d$, the message follows the path $e -> f -> g -> t$ to reach the final destination.

Figure \ref{fig:prop_rede2} shows a second case study, which represents a situation in which a minimum cut edge is selected for routing. Again a message is sent from source node $s$ to destination node $t$. In the original topology (graph $G$) Node $s$ has four alternatives to forward the message to: $a, c, g$, and $h$. However, function $Gamma$ is computed on $G'$ which is obtained by removing $s$ and all its adjacent edges from $G$. In  $G'$ the number of alternatives for node $s$ reduces to two nodes: $c$ and $a$. Function $Gamma$ returns a higher value for $c$, as through this node there are two disjoint paths to $t$. The message then reaches the destination through the path $c -> d -> t$. Note that this path has multiple alternatives that can be used in case there is some failure.

\subsection{Proof of Correctness}
\label{sec:prop_prova} 

In this section, we prove that FRR with backtracking succeeds routing a message from node $s$ to node $t$ across the network represented by graph $G$ if there is at least one working (non-failed) route from the source to the destination.

\textbf{Theorem 2.1.} Consider graph $G = (V,E)$ representing the network topology and that two arbitrary vertices $s,t \in V$, representing the source and the destination of a message $m$. Algorithm FRR with backtracking is able to successfully route a message from $s$ to $t$ if there is at least one working route from $s$ to $t$ in $G$.

\textbf{Proof.} A node running algorithm FRR with backtracking tries to send the message via the best-evaluated next hop of the corresponding entry of its routing table. The node corresponding to that next hop also does the same, and so on, until the message reaches $t$. However, if some node, say node $j$, of that route finds out that there is no working path to the destination, it will backtrack the message to the previous node from which it received the message, say node $i$. Node $i$ then tries the next best evaluated next hop, say $k$ and the process repeats: if there is no working route from node $k$ to $t$, it will return to node $i$. After node $i$ tries all alternatives without success, it backtracks the message. Backtracking can happen until the source, node $s$, which will also try all alternatives, and so if there is at least a single working route to the destination, that route will be found and used to successfully deliver the message to the destination $t$. $\Box$ 

\section{Simulation Results}
\label{cha:resultados}

This section presents experimental results obtained with simulation. The proposed algorithm was compared with Dijkstra’s shortest path algorithm \cite{dijkstra}. To be precise, the route computed by MaxFlowRouting (Algorithm \ref{fig:PseudoCodeGenRoutingTable}) was compared to the route computed by Dijkstra's algorithm. For the maximum flow evaluation, we employed the Highest-Label Preflow-Push algorithm \cite{preflow, preflow_networkx}. MaxFlowRouting was executed in the experiments considering three different pairs of weights for the Maximum Flow evaluation (MF) and for the Shortest Path evaluation (SP). The chosen sets were: 2 for MF and -5 for SP, which favors smaller route sizes; 5 for MF and -5 for SP, which balances the influence of both size and connectivity; and 5 for MF and -1 for SP, which favors route connectivity.

The two algorithms were compared using three metrics. The first metric is the average size of routes. The second metric is the sum of the degrees of all vertices along the routes. The idea is that a vertex with greater degree has a higher probability of having an alternative route to the destination. The third metric is the average number of backup routes available per vertex of each path. A backup route is a route that is disjoint from the original route selected by either of the algorithms. This metric is computed as follows. Initially, the edges of the original route are removed from the graph. Next, for each vertex of the route, except for the source and destination, we check whether there is a backup route to the destination. If there is indeed a backup route, the counter of backup routes of the corresponding vertex of the original route is incremented. In the end, the sum of all obtained backup routes is computed and it is divided by the size of the route, excluding the origin and destination vertices, thus obtaining the average number of disjoint backup routes per vertex. 

The routing algorithms were executed on multiple different topologies. For each graph, we computed the routes between all possible pairs of vertices. First, we generated Erdos-Renyi random graphs.  In those graphs the number of edges adjacent to each vertex is determined by a connectivity parameter $0 < C \leq 1$. The $C$ reflects the probability of any two processes being adjacent. Thus when $C = 1.0$, the connectivity is 100\%, that is, the network can be represented by a complete graph. Similarly, when $C=0.5$ the probability that there is an edge between any two processes is 50\%, and so on. In this paper, results are presented for simulations performed with the number of vertices equal to 100, 150, and 200 with connectivities $C = 0.1, 0.3, 0.5$, and $0.7$. Two other types of random graphs were also evaluated: the Barabási-Albert Preferential Attachment graph and the Watts-Strogatz Small World graph. These graphs were also generated with sizes equal to 100, 150, and 200 vertices.

Furthermore, we executed the routing algorithms also for real Internet topologies. We selected the most important topologies of Brazil, USA, Europe, and Japan: RNP \cite{rnp_topology}, Internet2 \cite{internet2_topology}, Géant \cite{geant_topology}, and Wide \cite{wide_topology}. For all metrics experiments, we only present results for the cases in which the route generated by MaxFlowRouting differed from that generated by Dijkstra’s algorithm. 

\subsection{Random Graphs with Varying Connectivity}

The Erdos-Renyi Random Graph \cite{random_networkx} is a graph with $N$ vertices and the probability that there an edge between any two vertices is $C$. Experiments were executed for graphs of sizes $N = 100, 150$, and $200$, and connectivity levels of $C = 0.1, 0.3, 0.5$ and $0.7$. The pairs of weights were: MF=2 and SP=-5, MF=5 and SP=-5, and MF=5 and SP=-1. The results are presented in table \ref{tab:random_graph}. 

\footnotesize
\begin{table*}
  \caption{Results for Erdos-Renyi random graphs with varying connectivity.}
  \label{tab:random_graph}
  \resizebox{\textwidth}{!}{
  \begin{tabular}{|ccc|ccc|c|ccc|}
    \toprule
    $N$ & $C$ & Weights & Avg size & Avg Deg Sum & Avg Backups & Route diff (\%) & D - Avg size & D - Avg Deg Sum & D - Avg Backups\\
    \midrule
    100 & 0.1 & MF=2, SP=-5 & 3.95 & 48.09 & 11.56 & 35.61\% & 3.54 & 37.45 & 8.11 \\
    100 & 0.1 & MF=5, SP=-5 & 4.27 & 54.70 & 12.43 & 38.52\% & 3.47 & 38.20 & 8.84 \\
    100 & 0.1 & MF=5, SP=-1 & 4.85 & 60.57 & 11.62 & 50.74\% & 3.47 & 36.33 & 8.20 \\
    150 & 0.1 & MF=2, SP=-5 & 3.85 & 68.59 & 17.81 & 36.38\% & 3.39 & 52.86 & 12.86 \\
    150 & 0.1 & MF=5, SP=-5 & 4.37 & 76.01 & 16.99 & 39.07\% & 3.38 & 50.59 & 12.36 \\
    150 & 0.1 & MF=5, SP=-1 & 4.24 & 69.67 & 17.26 & 44.06\% & 3.37 & 50.21 & 12.32 \\
    200 & 0.1 & MF=2, SP=-5 & 3.78 & 88.55 & 23.62 & 32.50\% & 3.25 & 67.12 & 17.50 \\
    200 & 0.1 & MF=5, SP=-5 & 4.32 & 99.99 & 23.00 & 40.51\% & 3.24 & 65.32 & 17.26 \\
    200 & 0.1 & MF=5, SP=-1 & 4.25 & 103.68 & 24.45 & 38.64\% & 3.20 & 67.29 & 17.92 \\
    \midrule
    100 & 0.3 & MF=2, SP=-5 & 3.03 & 98.60 & 33.67 & 22.80\% & 3.00 & 87.98 & 24.52 \\
    100 & 0.3 & MF=5, SP=-5 & 3.13 & 102.55 & 33.54 & 37.81\% & 3.00 & 89.64 & 25.77 \\
    100 & 0.3 & MF=5, SP=-1 & 4.14 & 106.73 & 34.59 & 21.25\% & 3.00 & 94.60 & 28.56 \\
    150 & 0.3 & MF=2, SP=-5 & 3.02 & 148.99 & 51.57 & 37.35\% & 3.00 & 137.32 & 41.03 \\
    150 & 0.3 & MF=5, SP=-5 & 3.09 & 150.45 & 50.44 & 32.33\% & 3.00 & 135.62 & 40.64 \\
    150 & 0.3 & MF=5, SP=-1 & 3.07 & 149.19 & 50.76 & 39.75\% & 3.00 & 133.99 & 39.87 \\
    200 & 0.3 & MF=2, SP=-5 & 3.09 & 196.16 & 65.21 & 41.71\% & 3.00 & 177.02 & 52.78 \\
    200 & 0.3 & MF=5, SP=-5 & 3.03 & 194.03 & 67.53 & 30.30\% & 3.00 & 177.59 & 53.70 \\
    200 & 0.3 & MF=5, SP=-1 & 3.07 & 194.70 & 66.17 & 30.51\% & 3.00 & 177.79 & 54.60 \\
    \midrule
    100 & 0.5 & MF=2, SP=-5 & 3.00 & 155.85 & 54.45 & 32.22\% & 3.00 & 142.10 & 40.97 \\
    100 & 0.5 & MF=5, SP=-5 & 3.01 & 163.87 & 56.80 & 16.32\% & 3.00 & 154.81 & 48.89 \\
    100 & 0.5 & MF=5, SP=-1 & 3.04 & 157.87 & 53.77 & 20.50\% & 3.00 & 145.92 & 44.28 \\
    150 & 0.5 & MF=2, SP=-5 & 3.01 & 234.37 & 80.28 & 23.95\% & 3.00 & 223.38 & 70.23 \\
    150 & 0.5 & MF=5, SP=-5 & 3.02 & 241.53 & 81.98 & 16.54\% & 3.00 & 232.04 & 74.12 \\
    150 & 0.5 & MF=5, SP=-1 & 3.02 & 230.44 & 78.18 & 21.77\% & 3.00 & 219.48 & 68.95 \\
    200 & 0.5 & MF=2, SP=-5 & 3.00 & 310.89 & 105.78 & 31.07\% & 3.00 & 298.18 & 93.42 \\
    200 & 0.5 & MF=5, SP=-5 & 3.01 & 313.88 & 106.80 & 11.96\% & 3.00 & 301.82 & 96.81 \\
    200 & 0.5 & MF=5, SP=-1 & 3.01 & 315.93 & 107.21 & 22.98\% & 3.00 & 302.64 & 95.56 \\
    \midrule
    100 & 0.7 & MF=2, SP=-5 & 3.00 & 213.45 & 72.62 & 16.30\% & 3.00 & 204.92 & 64.09 \\
    100 & 0.7 & MF=5, SP=-5 & 3.01 & 216.81 & 74.08 & 12.76\% & 3.00 & 199.06 & 57.44 \\
    100 & 0.7 & MF=5, SP=-1 & 3.00 & 214.21 & 72.28 & 9.27\% & 3.00 & 206.80 & 64.88 \\
    150 & 0.7 & MF=2, SP=-5 & 3.00 & 318.70 & 109.01 & 7.24\% & 3.00 & 307.75 & 98.06 \\
    150 & 0.7 & MF=5, SP=-5 & 3.00 & 318.62 & 107.48 & 24.80\% & 3.00 & 306.67 & 95.93 \\
    150 & 0.7 & MF=5, SP=-1 & 3.01 & 326.50 & 110.13 & 11.83\% & 3.00 & 318.72 & 103.31 \\
    200 & 0.7 & MF=2, SP=-5 & 3.00 & 432.45 & 146.78 & 14.04\% & 3.00 & 422.60 & 136.94 \\
    200 & 0.7 & MF=5, SP=-5 & 3.00 & 429.13 & 145.04 & 12.72\% & 3.00 & 418.94 & 135.21 \\
    200 & 0.7 & MF=5, SP=-1 & 3.00 & 428.44 & 145.06 & 15.18\% & 3.00 & 417.00 & 133.76 \\
    \bottomrule
  \end{tabular}
  }
\end{table*}
\normalsize

\begin{figure}[h]
  \centering
  \includegraphics[width=0.48\linewidth]{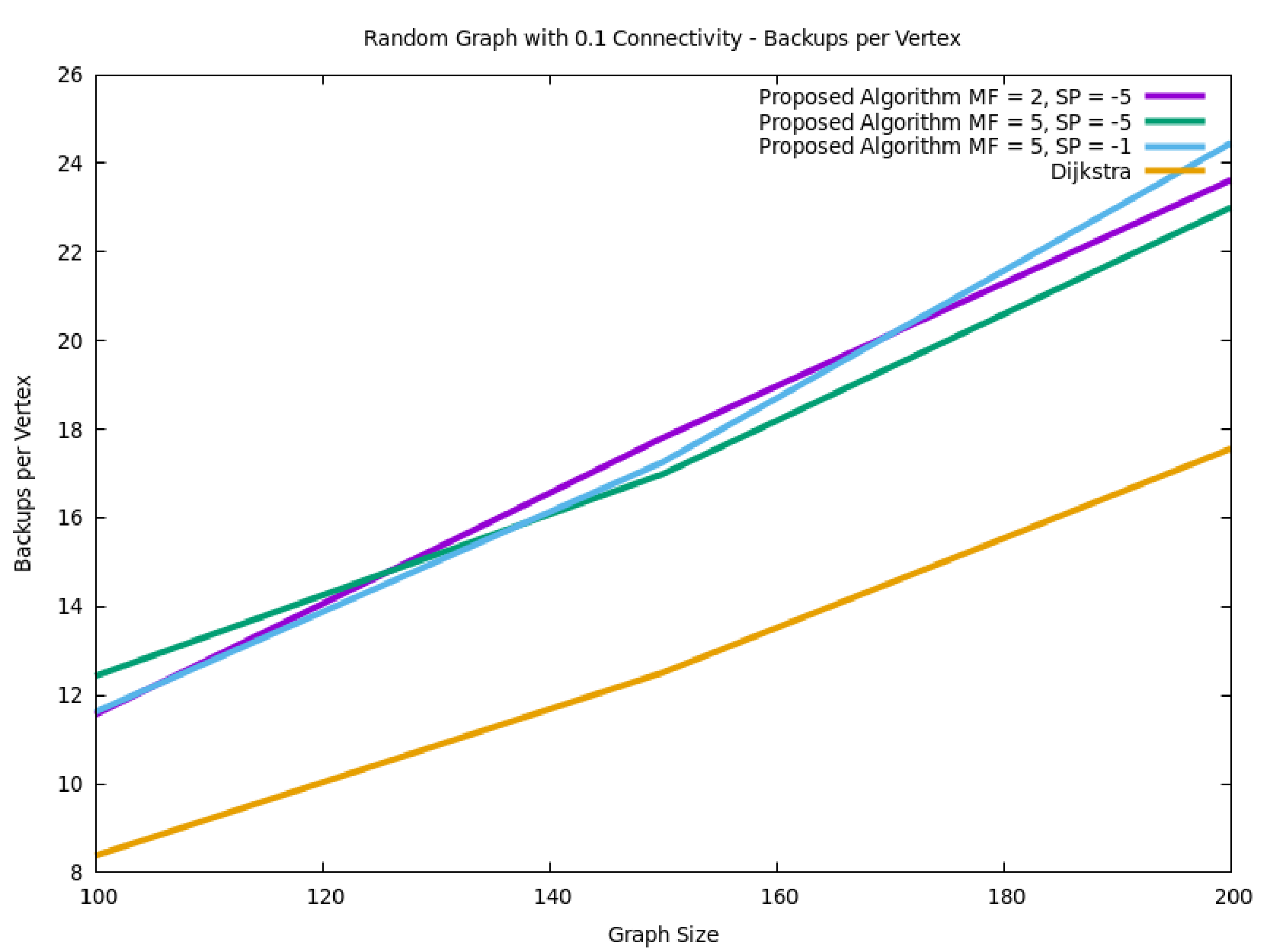}
  \includegraphics[width=0.48\linewidth]{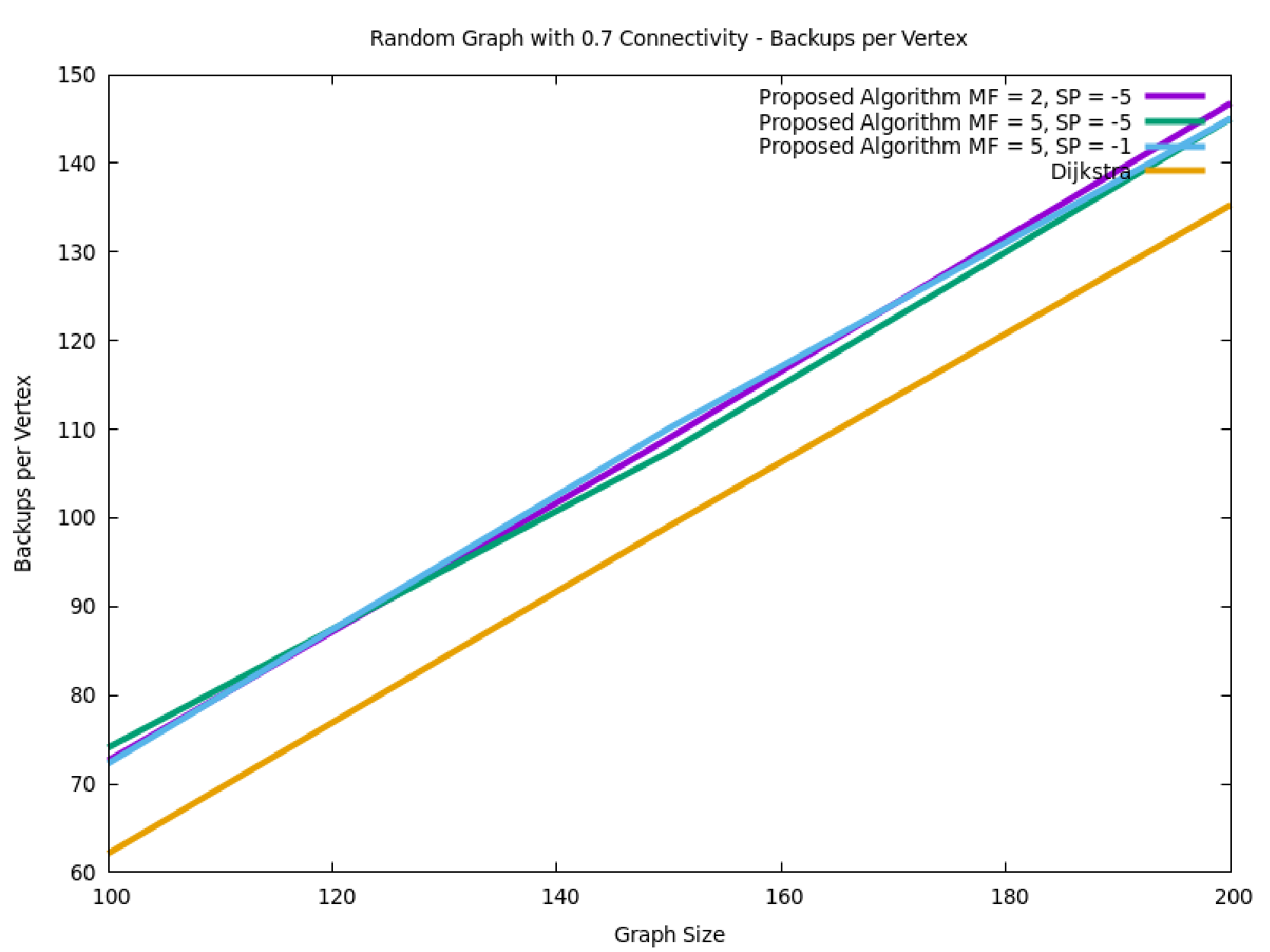}
  \caption{Comparison of the average backups per vertex computed by each algorithm, connectivity $C=0.1$ (left) and $C=0.7$ (right). Purple: MaxFlowRouting MF = 2, SP = -5. Green: MaxFlowRouting MF = 5, SP = -5. Blue: MaxFlowRouting with MF = 5, SP = -1. Yellow: Dijkstra.}
  \label{plt:random_backups}
\end{figure}

Figure \ref{plt:random_backups} shows clearly that MaxFlowRouting computed routes with more average backups per vertex than Dijkstra's algorithm. The number of backups per vertex increases in larger graphs. However, the difference between the two algorithm's results reduces as the graphs become more connected. It is also noticeable that the different pairs of weights had little effect on the proposed algorithm.

\begin{figure}[h]
  \centering
  \includegraphics[width=0.48\linewidth]{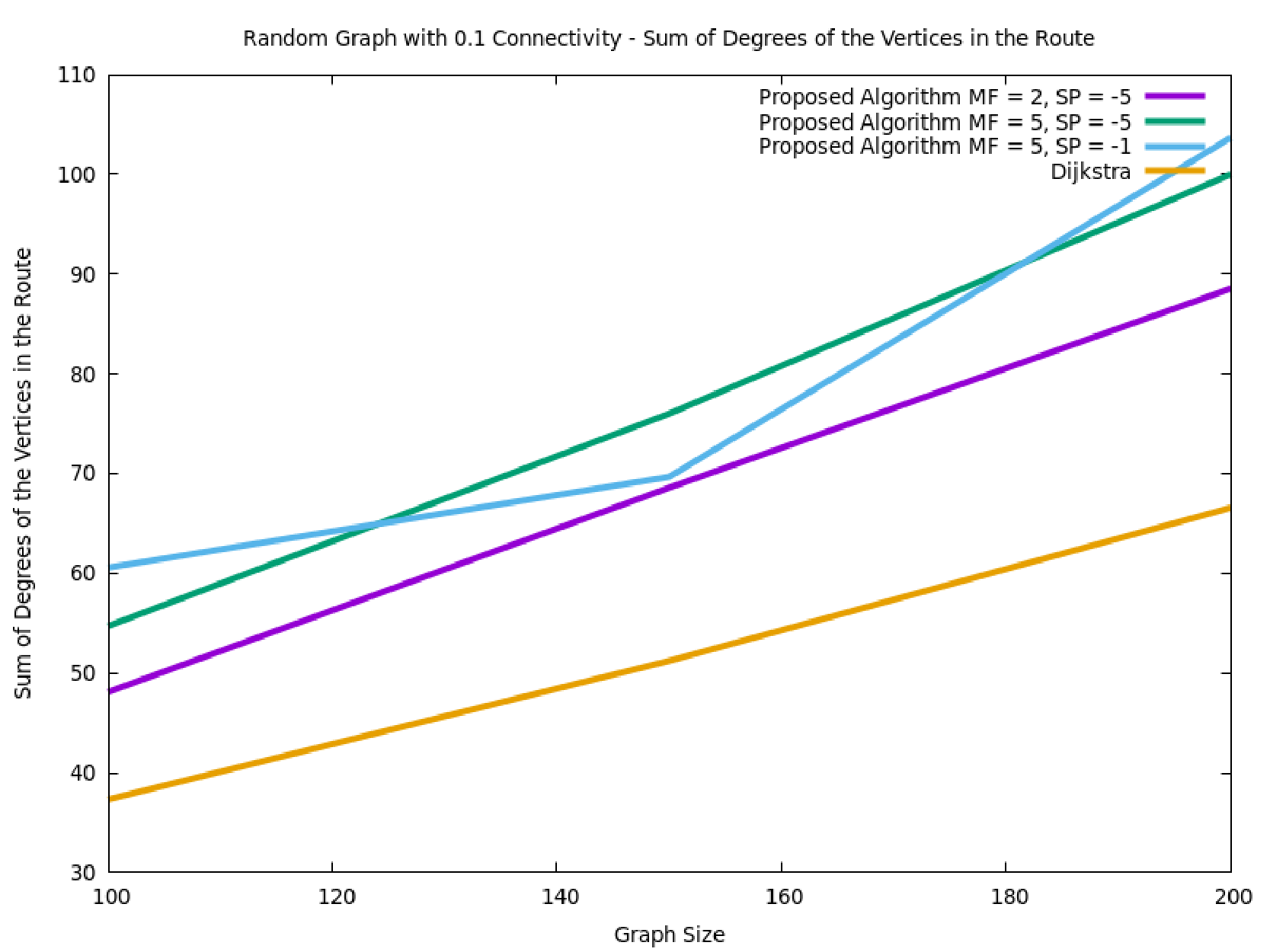}
  \includegraphics[width=0.48\linewidth]{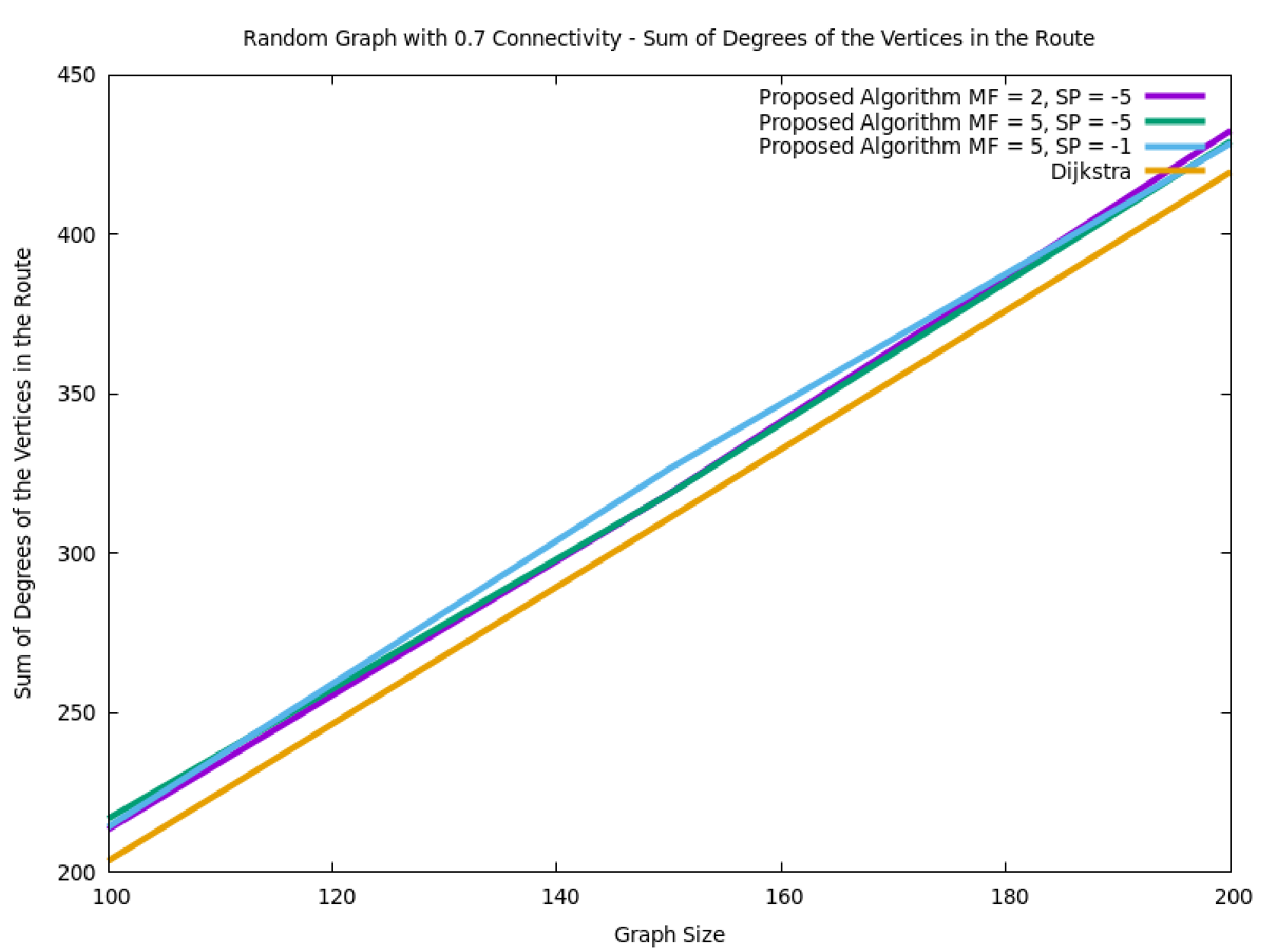}
  \caption{Comparison of the average vertex degrees computed by each algorithm, connectivity $C=0.1$ (left) and $C=0.7$ (right). Purple: MaxFlowRouting MF = 2, SP = -5. Green: MaxFlowRouting MF = 5, SP = -5. Blue: MaxFlowRouting with MF, and SP = -1. Yellow: Dijkstra.}
  \label{plt:random_degrees}
\end{figure}

Considering the average vertex degree metric, as shown in Figure \ref{plt:random_degrees} the number of backups per vertex of the routes computed by MaxFlowRouting is higher than that of Dijkstra's. The difference however reduces as the connectivity increases. The difference in the results given by the different sets of weights also reduces as the connectivity increases. Furthermore, the average vertex degrees increases with larger graphs.

\begin{figure}[h]
  \centering
  \includegraphics[width=0.48\linewidth]{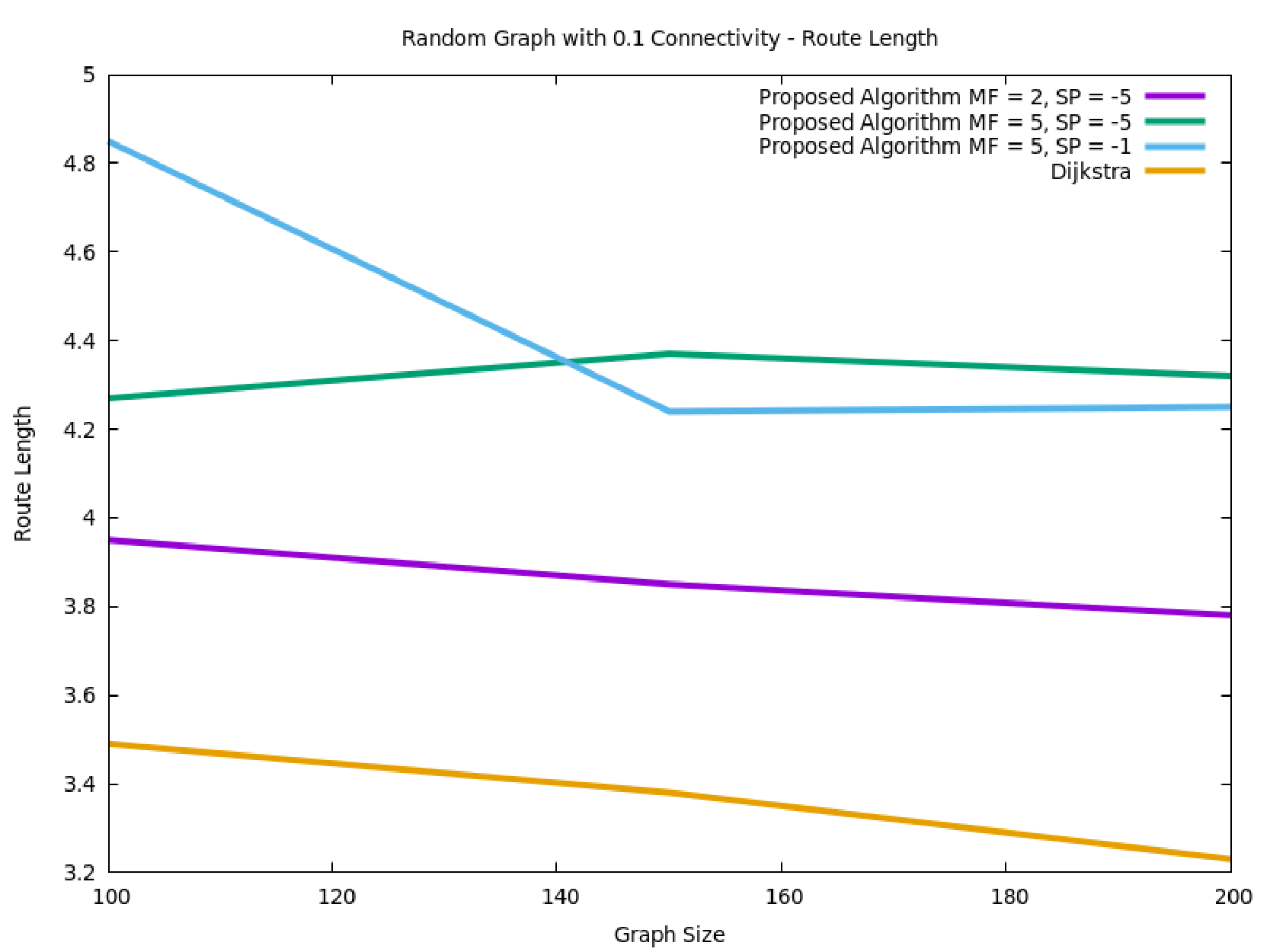}
  \includegraphics[width=0.48\linewidth]{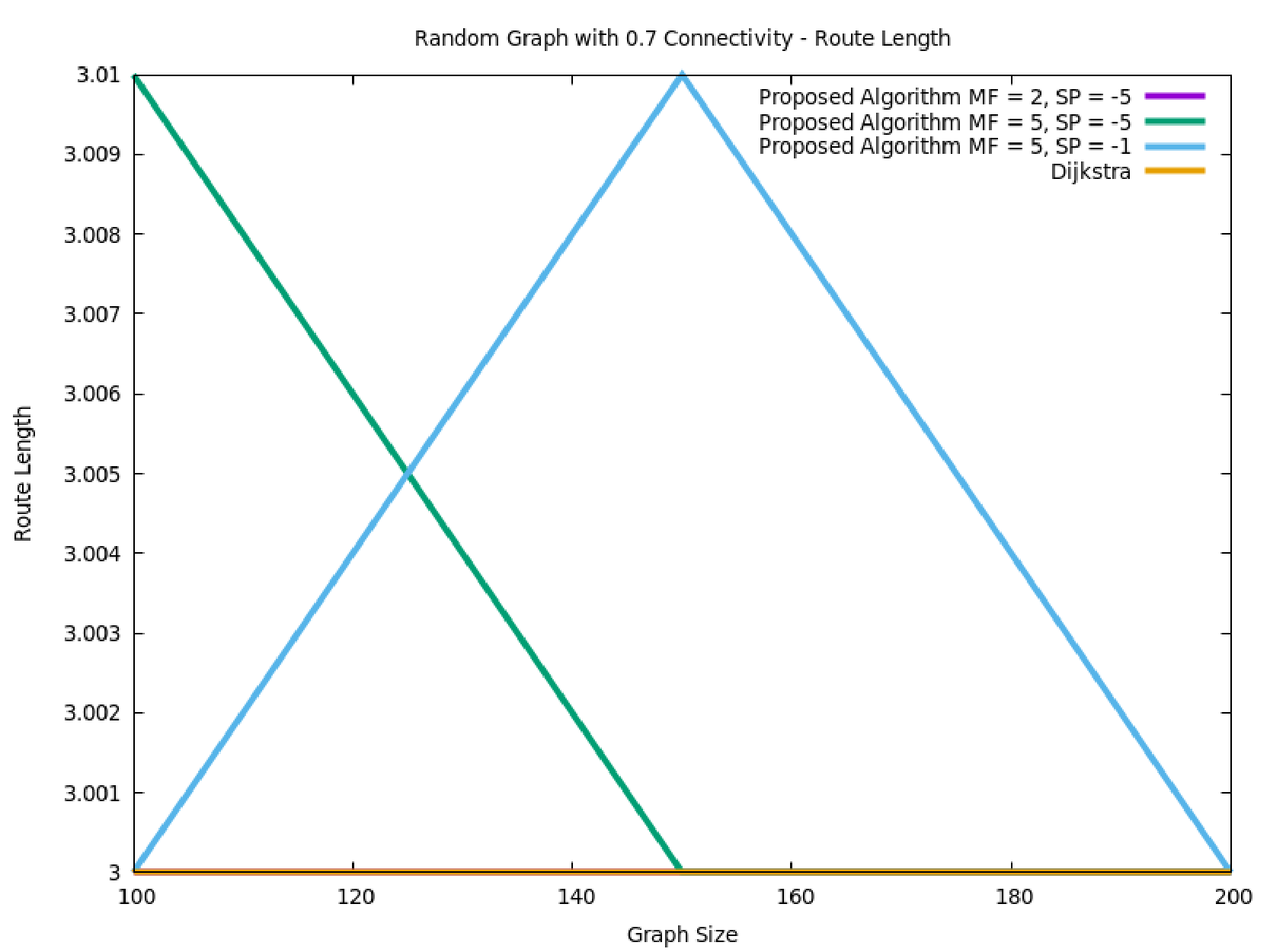}
  \caption{Comparison of the average route sizes computed by each algorithm, connectivity $C=0.1$ (left) and $C=0.7$ (right). Purple: MaxFlowRouting MF = 2 and SP = -5. Green: MaxFlowRouting MF = 5 and SP = -5. Blue: MaxFlowRouting MF = 5 and SP = -1. Yellow: Dijkstra.}
  \label{plt:random_size}
\end{figure}

The size of the routes computed by MaxFlowRouting was just slightly higher than those of Dijkstra's, as shown in Figure \ref{plt:random_size}. The difference decreased in graphs with higher connectivity. In the graph with connectivity 0.1, MaxFlowRouting with MF=2 and SP=-5 went from $3.95$ at $N=100$, to $3.85$ at $N=150$, and $3.78$ at $N=200$, while Dijkstra went from $3.54$ at $N=100$, $3.39$ at $N=150$ and $3.25$ at $N=200$. In the graph with connectivity 0.7, both algorithms had average route sizes of $3$, for all graph sizes ($N$).

\subsection{Preferential Attachment Graphs}

The Barabási-Albert Preferential Attachment Graph \cite{preferential_attachment_networkx} is a random graph with $N$ vertices. The graph is incrementally generated by adding vertices with degree $m$ edges, each of the edges connects the new vertex with existing vertices, preferentially with those of higher degrees, that is, that have more neighbors.
Experiments were executed for graphs of sizes $N = 100, 150$, and $200$, and vertices with $m = 3$ edges. tThe pairs of weights were: MF=2 and SP=-5, MF=5 and SP=-5, and MF=5 and SP=-1. Results are presented in table \ref{tab:preferential_attachment}. 

\footnotesize
\begin{table*}
  \caption{Results of the experiments with the Preferential Attachment Graphs.}
  \label{tab:preferential_attachment}
  \resizebox{\textwidth}{!}{
  \begin{tabular}{|cc|ccc|c|ccc|}
    \toprule
    $N$ & Weights & Avg size & Avg Deg Sum & Avg Backups & Route diff (\%) & D - Avg size & D - Avg Deg Sum & D - Avg Backups\\
    \midrule
    100 & MF=2, SP=-5 & 4.21 & 44.66 & 13.19 & 26.10\% & 4.10 & 36.93 & 9.80 \\
    100 & MF=5, SP=-5 & 4.27 & 49.61 & 15.18 & 29.11\% & 3.99 & 37.30 & 10.51 \\
    100 & MF=5, SP=-1 & 4.14 & 53.26 & 17.88 & 27.05\% & 3.98 & 43.10 & 13.52 \\
    \midrule
    150 & MF=2, SP=-5 & 4.26 & 55.27 & 17.64 & 25.13\% & 4.16 & 47.02 & 14.09 \\
    150 & MF=5, SP=-5 & 4.60 & 61.86 & 17.59 & 30.38\% & 4.16 & 45.50 & 13.29 \\
    150 & MF=5, SP=-1 & 4.38 & 69.67 & 22.94 & 27.69\% & 4.12 & 53.50 & 17.31 \\
    \midrule
    200 & MF=2, SP=-5 & 4.39 & 63.53 & 20.26 & 26.40\% & 4.30 & 55.02 & 16.86 \\
    200 & MF=5, SP=-5 & 4.60 & 69.65 & 20.74 & 31.14\% & 4.20 & 51.38 & 15.53 \\
    200 & MF=5, SP=-1 & 4.97 & 72.33 & 20.05 & 29.56\% & 4.23 & 52.78 & 16.08 \\
    \bottomrule
  \end{tabular}
  }
\end{table*}
\normalsize

\begin{figure}[h]
  \centering
  \includegraphics[width=\linewidth]{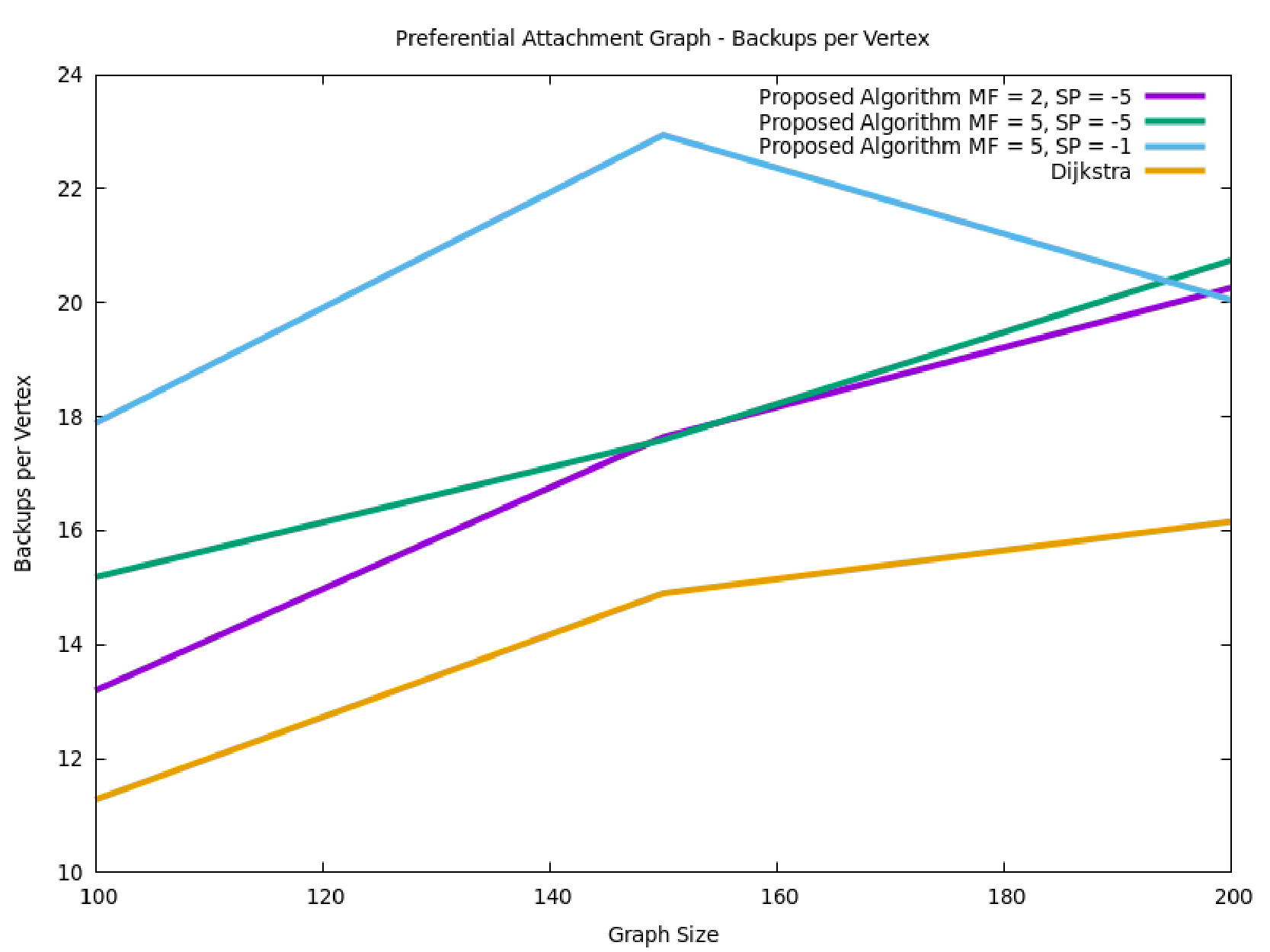}
  \caption{Comparison of the average number of backup routes per vertex. Purple: MaxFlowRouting MF = 2, SP = -5. Green: MaxFlowRouting MF = 5, SP = -5. Blue: MaxFlowRouting MF = 5, SP = -1. Yellow: Dijkstra.}
  \label{plt:preferential_backups}
\end{figure}

Figure \ref{plt:preferential_backups} shows that MaxFlowRouting produces routes with higher average backups per vertex than Dijkstra's algorithm with all 3 pairs of weights. However, the pair of weights MF = 5 and SP = -1 gave the proposed algorithm a high variability when $N$ varied, from 17.88 backup routes per vertex with $N=100$, going up to 22.94 backup routes per vertex with $N=150$, but dropping to 20.05 at $N=200$.

\begin{figure}[h]
  \centering
  \includegraphics[width=\linewidth]{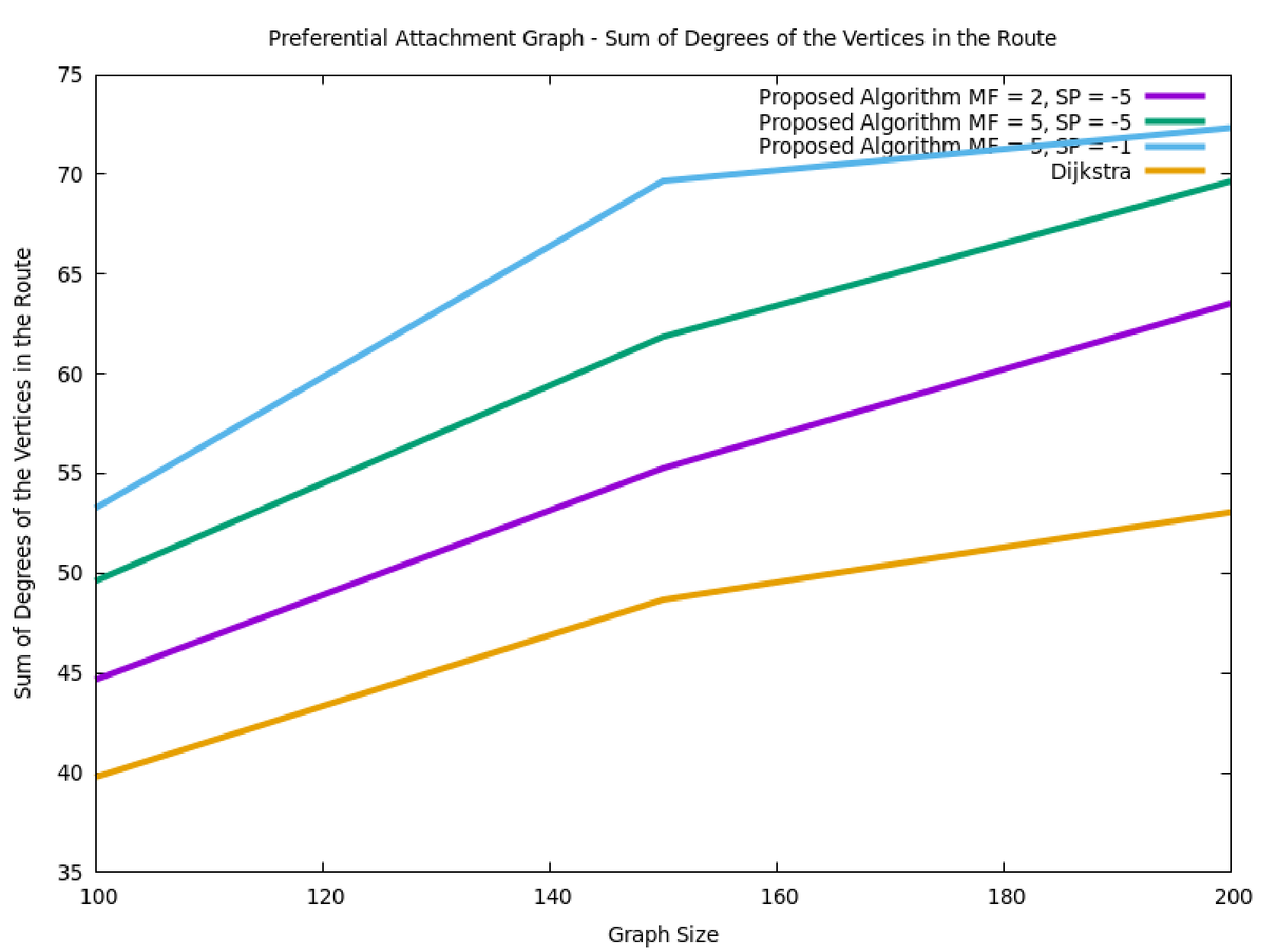}
  \caption{Comparison of the average route vertex degrees. Purple: MaxFlowRouting MF = 2, SP = -5. Green: MaxFlowRouting MF = 5, SP = -5. Blue: MaxFlowRouting MF = 5, SP = -1. Yellow: Dijkstra.}
  \label{plt:preferential_deg}
\end{figure}

Figure \ref{plt:preferential_deg} shows that the total vertex degrees of the routes computed by the proposed algorithm were higher than those of Dijkstra's. For all algorithms, the sum of vertex degrees grew as $N$ grew. MaxFlowRouting with the pair of weights MF = 5 and SP = -1 produced the routes with the highest total degrees.

\begin{figure}[h]
  \centering
  \includegraphics[width=\linewidth]{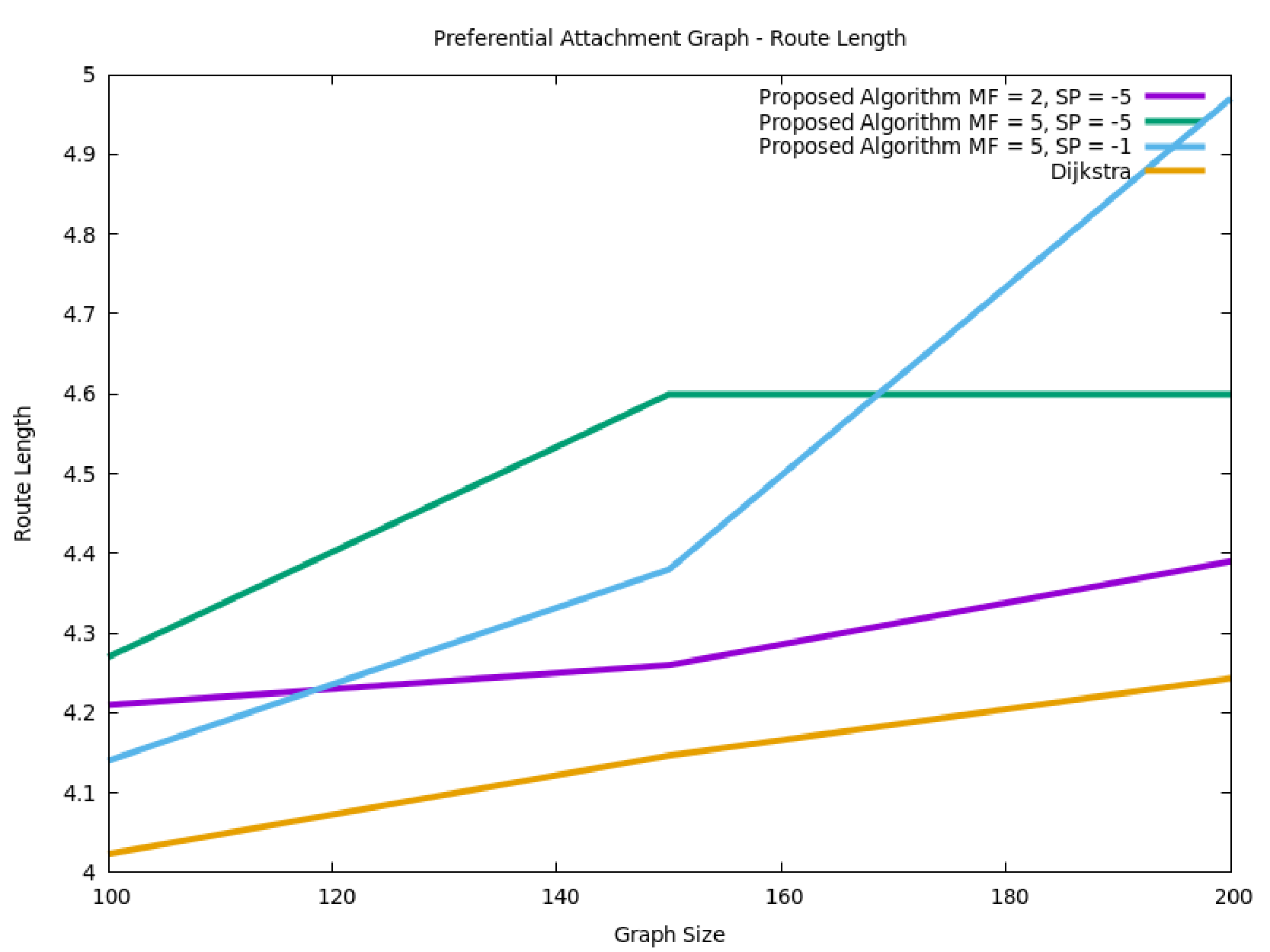}
  \caption{Comparison of the average route sizes. Purple: MaxFlowRouting MF = 2, SP = -5. Green: MaxFlowRouting MF = 5, SP = -5. Blue: MaxFlowRouting MF = 5, SP = -1. Yellow: Dijkstra.}
  \label{plt:preferential_size}
\end{figure}

Figure \ref{plt:preferential_size} shows that considering route size, the proposed algorithm with the pair of weights MF = 2 and SP = -5 produces routes that are just slightly longer than the routes produced by Dijkstra's algorithm. MaxFlowRouting with the weights MF = 5 and SP = -1 showed a significant increase in route sizes as $N$ grew.

\subsection{Small World Graphs}

The Watts-Strogatz Small World graph \cite{small_world_networkx} is a random graph with $N$ vertices generated initially as a ring-like topology in which each vertex is connected to its $k$ nearest neighbors. Then the graph is constructed by replacing with a probability $p$ each edge $(u,v)$ by a new edge $(u,t)$, where $t$ is a randomly selected vertex of the graph.

Experiments were executed on graphs of sizes $N = 100, 150$, and $200$, vertices with $k = 4$ neighbors, and the probability $p=0.4$ of edge replacement, and pairs of weights MF=2 and SP=-5, MF=5 and SP=-5, and MF=5 and SP=-1. 

\footnotesize
\begin{table*}
  \caption{Results of the experiments executed on Small World Graphs.}
  \label{tab:small_world}
  \resizebox{\textwidth}{!}{
  \begin{tabular}{|cc|ccc|c|ccc|}
    \toprule
    $N$ & Weights & Avg size & Avg Deg Sum & Avg Backups & Route diff (\%) & D - Avg size & D - Avg Deg Sum & D - Avg Backups\\
    \midrule
    100 & MF=2, SP=-5 & 5.61 & 26.41 & 3.06 & 25.39\% & 5.50 & 24.22 & 2.57 \\
    100 & MF=5, SP=-5 & 6.07 & 28.50 & 2.96 & 38.75\% & 5.04 & 21.66 & 2.34 \\
    100 & MF=5, SP=-1 & 10.76 & 54.91 & 3.37 & 60.28\% & 4.97 & 21.94 & 2.54 \\
    150 & MF=2, SP=-5 & 5.96 & 27.11 & 2.80 & 23.65\% & 5.86 & 25.29 & 2.43 \\
    150 & MF=5, SP=-5 & 6.47 & 31.77 & 3.22 & 37.15\% & 5.49 & 24.48 & 2.57 \\
    150 & MF=5, SP=-1 & 13.98 & 70.97 & 3.28 & 60.14\% & 5.27 & 23.00 & 2.46 \\
    200 & MF=2, SP=-5 & 6.16 & 28.57 & 2.91 & 22.23\% & 6.07 & 26.47 & 2.48 \\
    200 & MF=5, SP=-5 & 7.08 & 34.08 & 3.05 & 31.14\% & 5.82 & 25.45 & 2.45 \\
    200 & MF=5, SP=-1 & 15.46 & 74.51 & 2.94 & 70.51\% & 5.64 & 24.17 & 2.36 \\
    \bottomrule
  \end{tabular}
  }
\end{table*}
\normalsize

\begin{figure}[h]
  \centering
  \includegraphics[width=\linewidth]{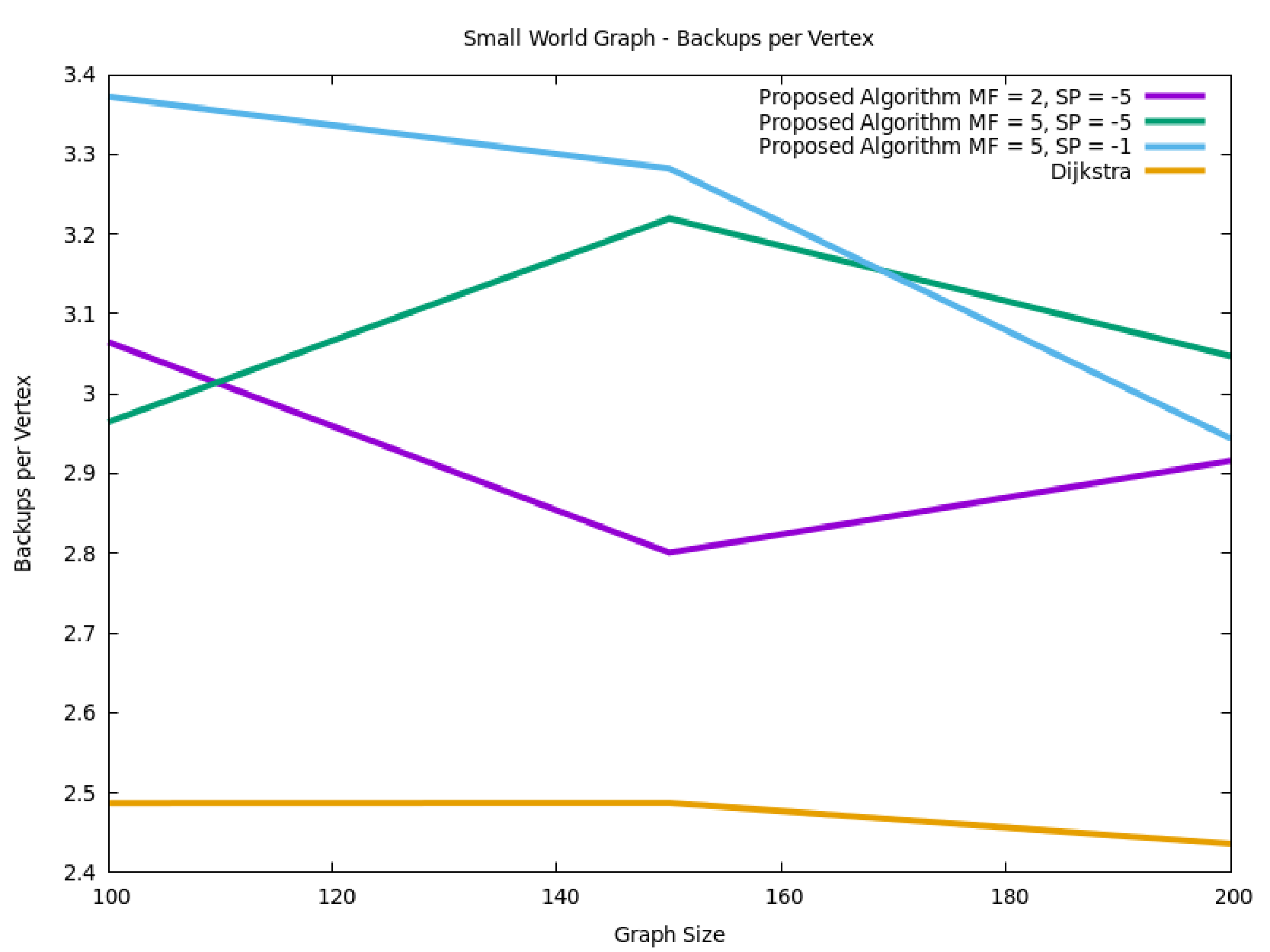}
  \caption{Comparison of the average backups per vertex. Purple: MaxFlowRouting MF = 2, SP = -5. Green: MaxFlowRouting MF = 5, SP = -5. Blue: MaxFlowRouting MF = 5 and SP = -1. Yellow: Dijkstra.}
  \label{plt:small_world_backups}
\end{figure}

Figure \ref{plt:small_world_backups} shows the average number of backup routes per vertex of the routes computed by Dijkstra’s and the proposed algorithm. It is noticeable that the proposed algorithm surpassed Dijkstra’s algorithm for all three pairs weights employed. However, it is also possible to see that the variation was quite high as the graph sizes ($N$) varied. Routes computed with MaxFlowRouting with the weights MF=5 and SP=-1 presented 3.37 backups per vertex with $N=100$, dropped slightly to 3.28 with $N=150$ and even further to 2.94 with $N=200$. Dijkstra was surpassed on larger graphs by the proposed algorithm with the weights MF=5 and SP=-5, which had 2.96 backups per vertex at $N=100$, going up to $3.22$ backups per route at $N = 150$ and dropping to $3.04$ with $N = 200$. 

\begin{figure}[h]
  \centering
  \includegraphics[width=\linewidth]{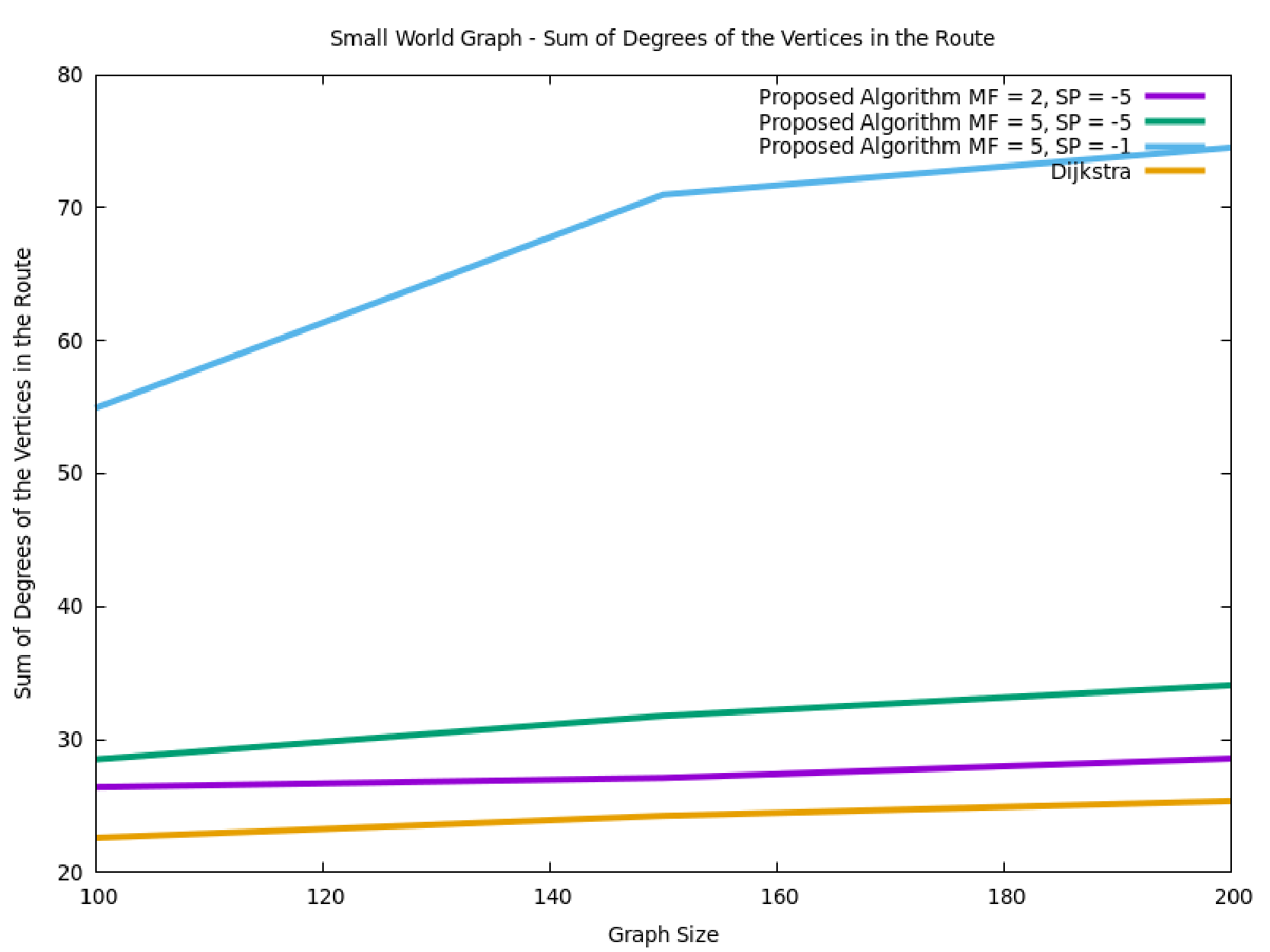}
  \caption{Comparison of the average sum of route vertices degrees. Purple: MaxFlowRouting MF = 2, SP = -5. Green: MaxFlowRouting MF = 5, SP = -5. Blue: MaxFlowRouting MF = 5, SP = -1. Yellow: Dijkstra.}
  \label{plt:small_world_deg}
\end{figure}

Figure \ref{plt:small_world_deg} shows that the largest average sum of vertex degrees was for routes computed by MaxFlowRouting with the pair of weights MF=5 and SP=-1. This was expected, as this pair of weights favors route connectivity. With other pairs of weights, MaxFlowRouting also showed an advantage compared to Dijkstra’s algorithm considering this metric.

\begin{figure}[h]
  \centering
  \includegraphics[width=\linewidth]{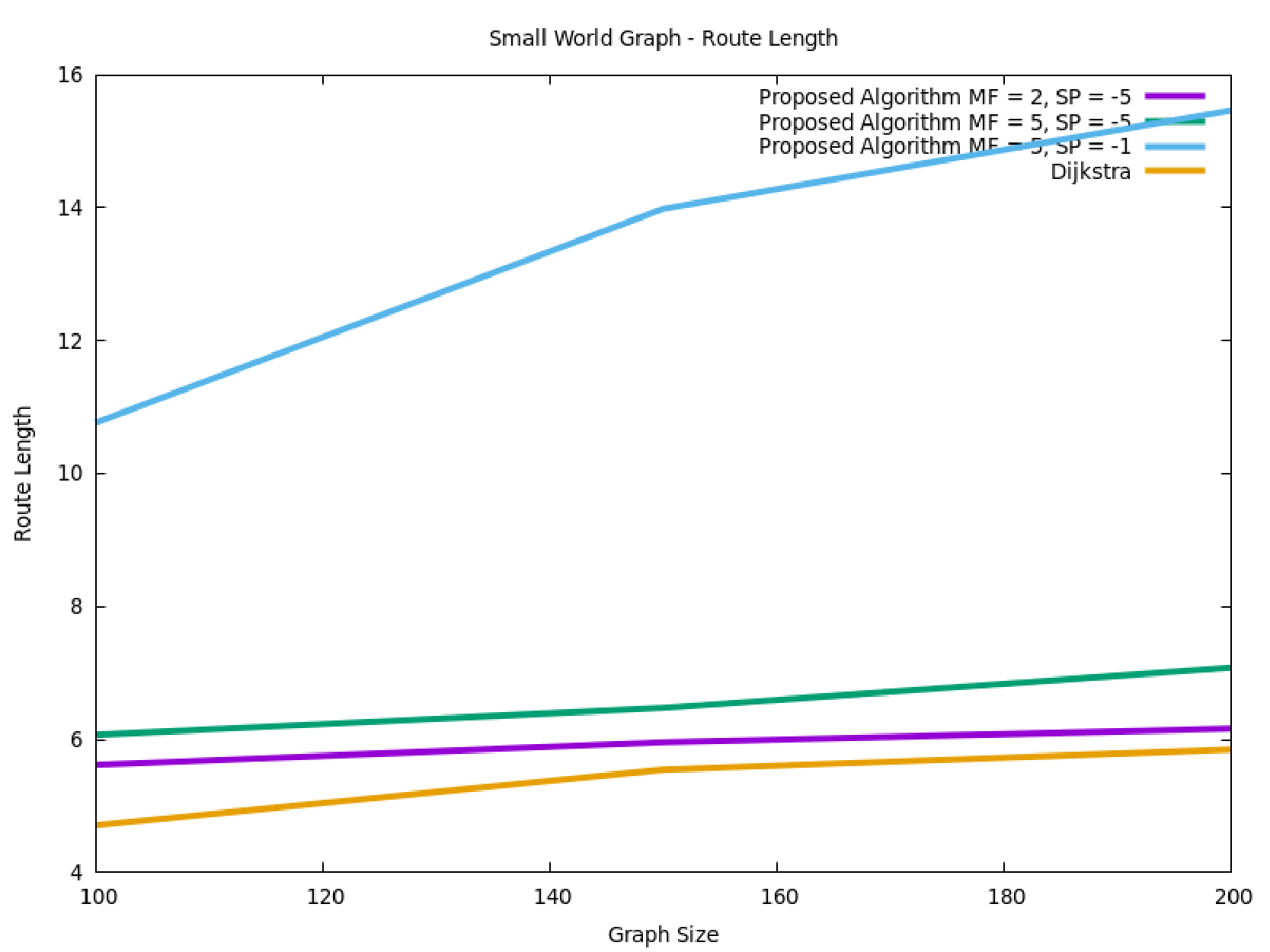}
  \caption{Comparison of the average route sizes. Purple: MaxFlowRouting MF = 2, SP = -5. Green: MaxFlowRouting MF = 5, SP = -5. Blue: MaxFlowRouting MF = 5, SP = -1. Yellow: Dijkstra.}
  \label{plt:small_world_size}
\end{figure}

In terms of average route size, Figure \ref{plt:small_world_size} shows that the routes computed by MaxFlowRouting with weights MF=2 and SP=-5 were the shortest, compared to those computed with the other two pairs of weights (MF=5 and SP=-5 and MF=5 and SP=-1). In the best case, the routes computed with the proposed algorithm were just slightly larger than Dijkstra’s, which computes the routes with the shortest sizes possible.

\subsection{Internet Topologies: RNP, Internet2, Géant, and Wide}

Experiments were also conducted on graphs representing real Internet topologies from the U.S.A., Europe, Brazil, and Japan. The Internet2 backbone \cite{internet2_topology} has a topology with access points all around the U.S.A. The GÉANT network \cite{geant_topology} is a European backbone that connects research and educational networks around the continent. The RNP Ipê network \cite{rnp_topology} is a Brazilian academic network that covers the whole country. The Wide network \cite{wide_topology} is a backbone that connects universities, research institutions, and companies in Japan. The graph representing the Internet2 topology has $N = 54$ vertices, the Géant topology has $N=44$ vertices, the RNP topology has $N = 28$ vertices, and the Wide topology has $N = 14$ vertices. The results are shown in table \ref{tab:topologies}.

\begin{table*}
  \caption{Results observed in experiments with the Intenet Topologies}
  \label{tab:topologies}
  \resizebox{\textwidth}{!}{
  \begin{tabular}{|cc|ccc|c|ccc|}
    \toprule
    Topology & Weights & Avg size & Avg Deg Sum & Avg Backups & Route diff (\%) & D - Avg size & D - Avg Deg Sum & D - Avg Backups\\
    \midrule
    Internet2 & MF=2, SP=-5 & 9.72 & 26.76 & 0.83 & 12.30\% & 9.72 & 25.55 & 0.63 \\
    Internet2 & MF=5, SP=-5 & 9.49 & 26.38 & 0.87 & 16.35\% & 9.23 & 24.50 & 0.67 \\
    Internet2 & MF=5, SP=-1 & 10.42 & 28.70 & 0.83 & 44.09\% & 9.02 & 24.18 & 0.73 \\
    Géant & MF=2, SP=-5 & 6.46 & 25.16 & 2.32 & 25.31\% & 6.46 & 25.04 & 2.32 \\
    Géant & MF=5, SP=-5 & 6.77 & 26.42 & 2.36 & 34.20\% & 6.41 & 23.77 & 2.08 \\
    Géant & MF=5, SP=-1 & 7.61 & 30.11 & 2.37 & 43.09\% & 6.25 & 23.00 & 2.01 \\
    RNP & MF=2, SP=-5 & 4.44 & 25.79 & 5.73 & 17.99\% & 4.44 & 22.17 & 4.11 \\
    RNP & MF=5, SP=-5 & 4.42 & 25.54 & 5.71 & 19.31\% & 4.35 & 21.26 & 3.85 \\
    RNP & MF=5, SP=-1 & 4.74 & 27.02 & 5.57 & 19.84\% & 4.38 & 21.32 & 3.84 \\
    WIDE & MF=2, SP=-5 & 3.66 & 17.66 & 5.00 & 3.29\% & 3.66 & 13.66 & 2.33 \\
    WIDE & MF=5, SP=-5 & 3.83 & 18.83 & 5.00 & 6.59\% & 3.33 & 11.50 & 1.83 \\
    WIDE & MF=5, SP=-1 & 3.83 & 18.83 & 5.00 & 6.59\% & 3.33 & 11.50 & 1.83 \\
    \bottomrule
  \end{tabular}
  }
\end{table*}

\begin{figure}[h]
  \centering
  \includegraphics[width=\linewidth]{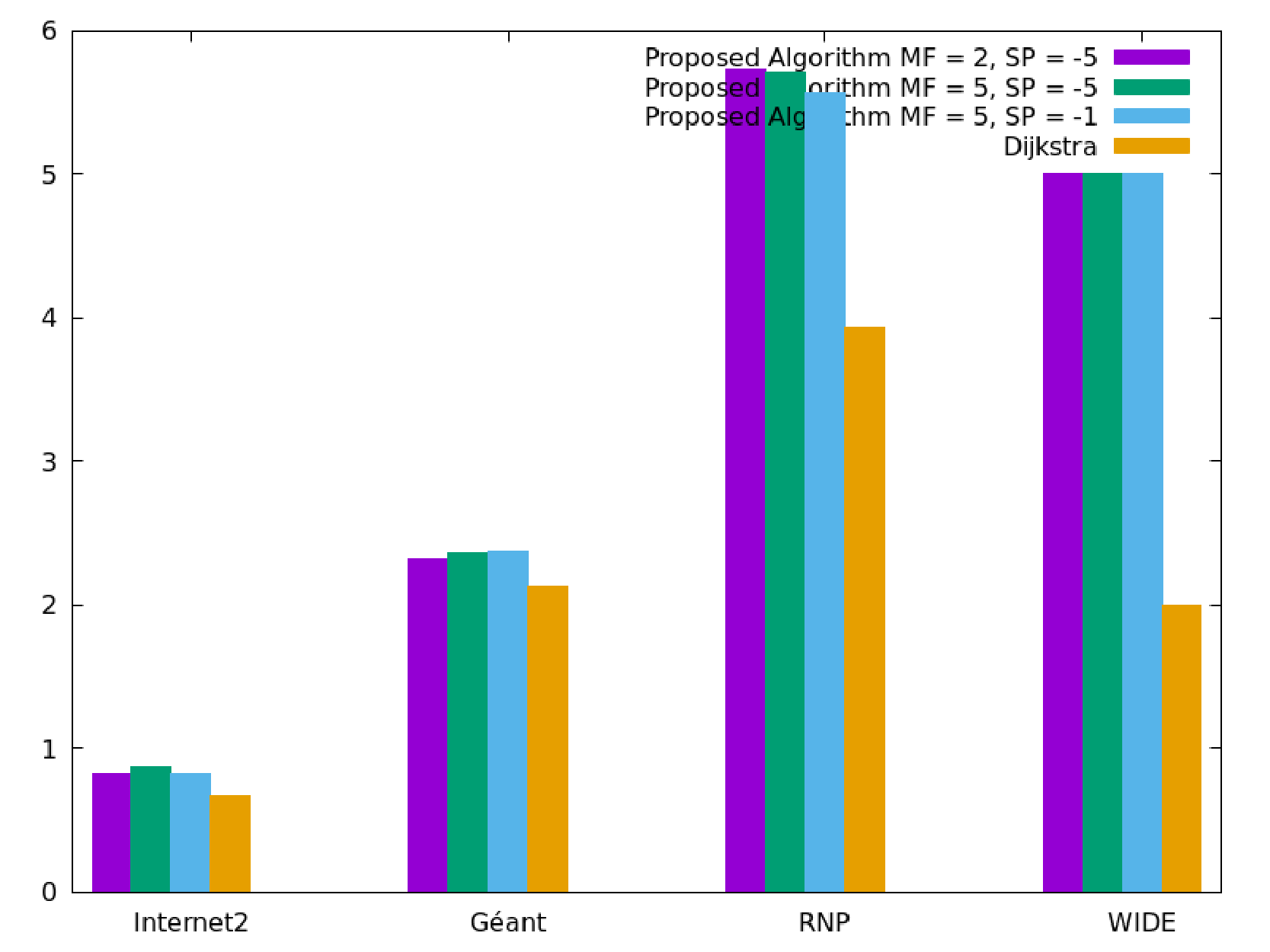}
  \caption{Comparison of the average backups per vertex for real Internet backbone topologies. Purple: MaxFlowRouting MF = 2, SP = -5. Green: MaxFlowRouting MF = 5, SP = -5. Blue: MaxFlowRouting MF = 5, SP = -1. Yellow: Dijkstra. From left to right: Internet2, Géant, RNP, and Wide topologies.}
  \label{plt:topologies_backups}
\end{figure}

Figure \ref{plt:topologies_backups} shows that MaxFlowRouting with all 3 pairs of weights, produces routes with more backups per vertex than Dijkstra's algorithm for all networks. It is also possible to see that there was little variation between the results produced by the proposed algorithm with the different pairs of weights. The average number of backups per vertex did vary from one network to another.

\begin{figure}[h]
  \centering
  \includegraphics[width=\linewidth]{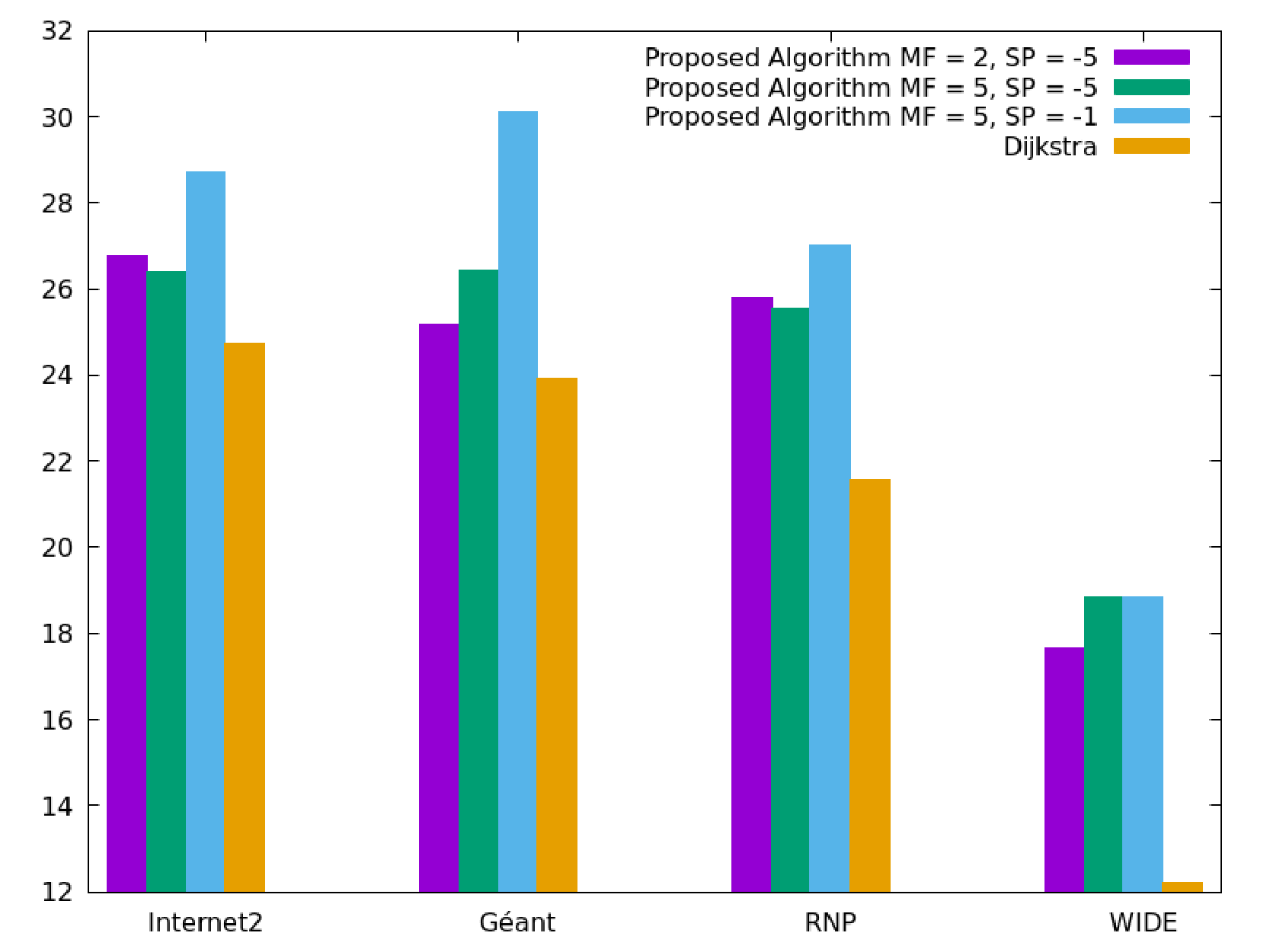}
  \caption{Comparison of the average sum of route vertex degrees. Purple: MaxFlowRouting MF = 2, SP = -5. Green: MaxFlowRouting MF = 5, SP = -5. Blue: MaxFlowRouting MF = 5, SP = -1. Yellow: Dijkstra. From left to right: Internet2, Géant, RNP, and Wide topologies.}
  \label{plt:topologies_degrees}
\end{figure}

As shown in Figure \ref{plt:topologies_degrees}, MaxFlowRouting produced the routes with the largest average sum of vertex degrees for pair of weights MF = 5 and SP = -1. But even the other pairs of weights computed routes with average sums of vertex degrees that surpassed Dijkstra's, with a particularly large gap for the Wide topology.

\begin{figure}[h]
  \centering
  \includegraphics[width=\linewidth]{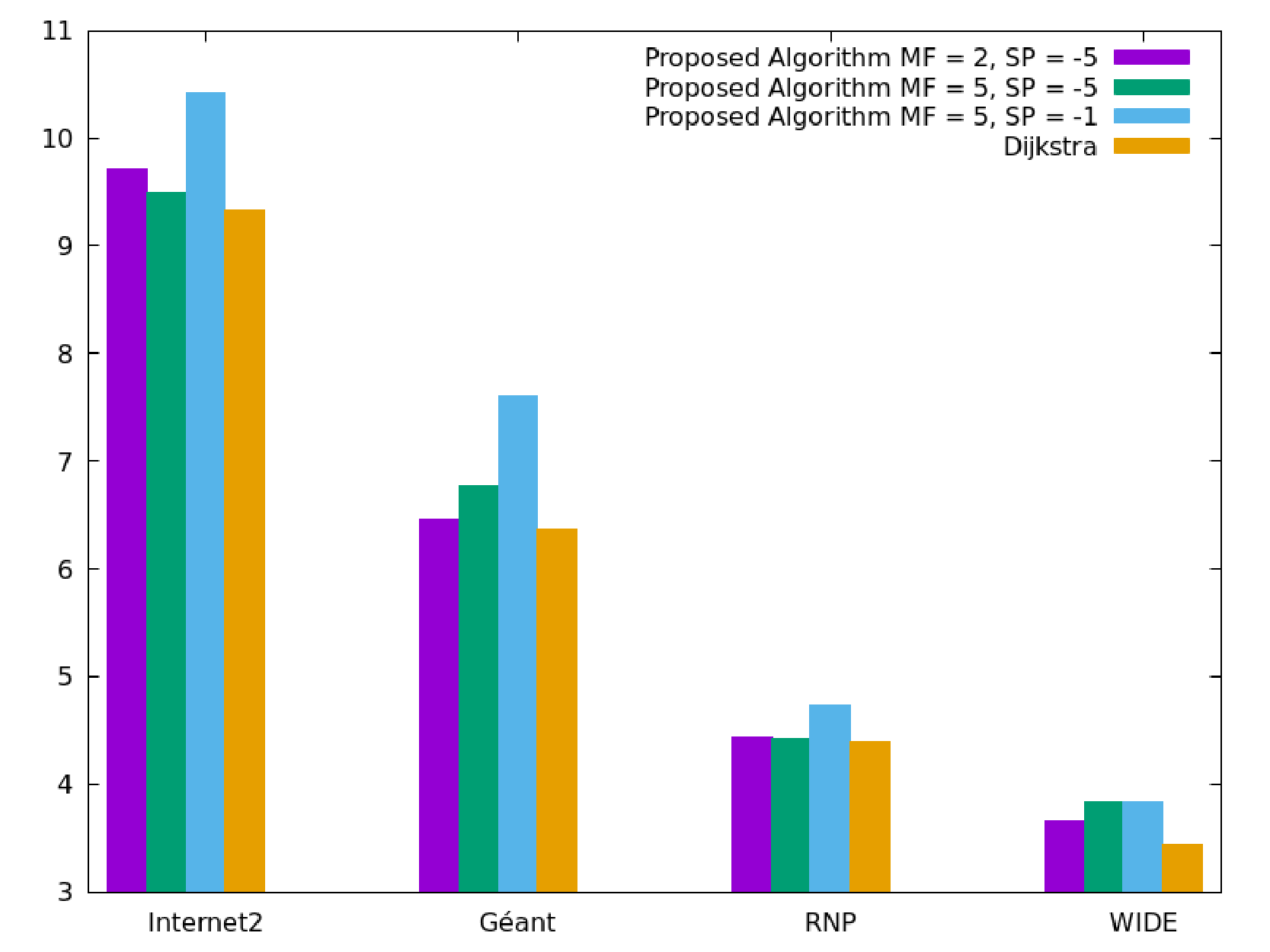}
  \caption{Comparison of the average route sizes. Purple: MaxFlowRouting MF = 2, SP = -5. Green: MaxFlowRouting MF = 5, SP = -5. Blue: MaxFlowRouting MF = 5, SP = -1. Yellow: Dijkstra. From left to right: Internet2, Géant, RNP, and Wide topologies.}
  \label{plt:topologies_lenght}
\end{figure}

Figure \ref{plt:topologies_lenght} shows that the routes computed by the proposed algorithm with the pair of weights MF = 2 and SP = -5 were the shortest, compared to those produced with the other pairs of weights. The routes produced by it were slightly larger than the ones produced by Dijkstra's algorithm, which computes the shortest paths. For the Internet2 topology, the pair of weights MF = 5 and SP = -5  outperformed the other pairs of weights considering route sizes and real Internet backbone topologies.

\section{Related Work}
\label{sec:rot_trab_relacionados}

IP FRR (IP Fast Reroute) \cite{gjoka2007evaluation} was first proposed in the context of the convergence issues of BGP (Border Gateway Protocol), that were very evident at that time \cite{duarte2004delivering}. With the increasing demand for low-latency by Internet applications, the slow convergence translates into an important concern. 
One of the first IPFRR strategies for rerouting traffic through pre-computed alternatives was proposed by Kvalbein and others \cite{mrc}. The central idea of the strategy, called \textsl{Multiple Routing Configurations (MRC)}, consists of using the known topology of the network and building a set of routing alternatives. MRC was further improved by Tarik \textsl{et. al.}.

The work by Cicik and others \cite{cicik2009} is an alternative to IPFRR (IP Fast ReRoute) that maintains several competing views of the topology of the network, allowing rapid recovery after failures occur. Another approach that proved to be effective for fault-tolerant routing on the Internet and which consists of taking direct action on the IP protocol itself, is the use of not-via addresses \cite{ipnotvia2009}. Not-via is a mechanism for protecting the IP routing against failed links or routers. Routers learn about not-via addresses so that they can avoid certain paths. 

After a link or router fails, routing protocols present a latency to update the routing tables so that they reflect the new topology. There are algorithms for discovering and maintaining the topology of dynamic networks, such as \cite{nassu2007topology}. Distributed diagnosis algorithms for general topology networks are also used to monitor the network and disseminate new event information that can be employed to update the routing tables \cite{duarte2003distributed, duarte1997non}. There are multiple distributed strategies for monitoring the network topology, including those based on evolutionary algorithms \cite{nassu2005comparison}. Intelligent strategies have also been explored for communications in dynamic networks \cite{banzi2011approach}. Although traditional diagnosis algorithms require perfect tests, recently a diagnosis model has been presented for asynchronous systems, which is based on imperfect tests \cite{duarte2023missing}.

Ohara and others presents in \cite{ohara2009} a new family of algorithms that, given the network topology, calculates a directed acyclic graph (DAG - Directed Acyclic Graph) that also takes into account the maximum flow. The efficiency of the proposed method is evaluated through the complexity of the developed algorithms and also through the simulation of the strategy on different topologies of Internet ASes (Autonomous Systems). 

Although MaxFlowRouting is related to the proposal in \cite{schroeder2007fault}, as both use maximum flow evaluation for robust routing, the main difference is that routing algorithm based on maximum flow evaluation is presented that does not require previous knowledge of the topology and can be applied to external routing. Routes are computed as a packet reaches a router given the path traversed so far and the alternatives to the destination. The algorithm disregards the routes that go through any node already visited by the message when computing the maximum flow. In this way, this process of routing is dependent on the context of \textsl{each} message, and also on the set of nodes visited by each message at each node traveled. That algorithm presents a high cost, which prevents its practical application to real networks.

Yet another related approach for fault-tolerant routing is the set of connectivity criteria proposed by Cohen and others \cite{cohen2011connectivity}. The set of quantitative criteria assesses node connectivity. Each network node receives a label -- called connectivity number -- that indicates the number of edges that have to fail for that node to disconnect from the network. The authors show that the connectivity numbers can be computed in polynomial time. Connectivity numbers have been employed to choose a so called ``routing proxy'' \cite{duarte1999formal} which is a relay node on the application level that can forward a message to a given destination. The path through the routing proxy to the destination has been called a ``detour'', and has employed in the context of network management \cite{cohen2001netmgmt} and general application PDUs \cite{cohen2001fault}.

As mentioned before maximum flow is equivalent to finding the minimum cut. A cut tree \cite{cohen2017parallel} is a combinatorial structure that represents the edge-connectivity between all pairs of nodes, solving the all pairs minimum cut problem efficiently. Parallel cut algorithms have been proposed to speed up the computation of a cut tree of a given graph, which can be employed to compute the minimum cut between all pairs of nodes and thus speedup the computation of routes by MaxFlowRouting. Parallel versions of both the Gusfield algorithm \cite{cohen2011parallel} and the Gomory-Hu algorithm \cite{cohen2012parallel, maske2020speeding} have been proposed.

Path diversity has been considered in multiple environments. Chen and others \cite{chen2016path} propose a fault-tolerant routing algorithm for a NoC (Network-on-Chip). Besides fault-resilient the authors also consider traffic balancing through the multiple routes. Hasan and others \cite{hasan2017optimizing} propose a particle multi-swarm optimization (PMSO) routing algorithm to construct, recover, and select k-disjoint paths that tolerate faults while satisfying quality of service requirements in the context of the Internet of Things. Finally, Shyama and others \cite{shyama2022self} propose a fault-tolerant routing algorithm based on swarm intelligence that relies on multiple paths to the destination in the context of wireless sensor networks, which require metrics such as minimizing energy consumption.

\section{Conclusions}
\label{cha:conclusao}

This work proposed MaxFlowRouting, an algorithm that combines maximum flow evaluation with route size to compute robust routes. The purpose is to use MaxFlowRouting in the context of FRR, as routes are well connected and present more alternatives to reach a destination in case of a route failure. Besides specifying and evaluating MaxFlowRouting, we also specify a FRR algorithm with backtracking that we prove to able to successfully forward a packet from the source to the destination, if there is at least one fully correct route connecting those nodes. Experimental results were also presented comparing MaxFlowRouting and Dijkstra's shortest path algorithm on several topologies. Results show that MaxFlowRouting produces robust routes that consist of nodes with larger degrees, provide a larger number of backup routes, and the route sizes are just slightly larger than those produced by Dijkstra's.

Future work should extend MaxFlowRouting to use of other criteria, such as traffic load and quality of service parameters in addition to the maximum flow and distance to compute the routes. An experimental evaluation of FRR executed with Dijkstra and MaxFlowRouting, as well of the impact of the backup routes provided by both algorithms is also left as future work. The construction of a robust version of the OSPF (Open Shortest Path First) protocol based on MaxFlowRouting is also foreseen as future work.

\bibliographystyle{ieeetr}
\bibliography{main}

\end{document}